\documentclass[acmsmall]{acmart}
\usepackage{romannum}
\usepackage{enumitem}
\usepackage{multirow}
\usepackage{subfigure}
\usepackage{bbm}
\usepackage{makecell}
\usepackage{abraces}
\usepackage[ruled]{algorithm2e}
\usepackage{graphicx}
\usepackage{amsthm}
\usepackage{xcolor}
\newtheorem{definition}{Definition}
\newtheorem{theorem}{Theorem}
\newtheorem{proposition}[theorem]{Proposition}
\DeclareMathOperator{\Tr}{Tr}
\AtBeginDocument{%
  }

\setcopyright{acmcopyright}
\copyrightyear{2018}
\acmYear{2018}
\acmDOI{XXXXXXX.XXXXXXX}

\acmJournal{JACM}
\acmVolume{37}
\acmNumber{4}
\acmArticle{111}
\acmMonth{8}

\begin{document}
\pagenumbering{arabic}
\title{Balancing Embedding Spectrum for Recommendation}

\author{Shaowen Peng}
\email{peng.shaowen@naist.ac.jp}
\affiliation{%
  \institution{NARA Institute of Science and Technology}
  \city{Nara}
  \country{Japan}
}

\author{Kazunari Sugiyama}
\email{sugiyama-k@g.osaka-seikei.ac.jp}
\affiliation{%
  \institution{Osaka Seikei University}
  \city{Osaka}
  \country{Japan}
}

\author{Xin Liu}
\email{xin.liu@aist.go.jp}
\affiliation{%
  \institution{National Institute of Advanced Industrial Science and Technology}
  \city{Tokyo}
  \country{Japan}
}

\author{Tsunenori Mine}
\email{mine@m.ait.kyushu-u.ac.jp}
\affiliation{%
  \institution{Kyushu University}
  \city{Fukuoka}
  \country{Japan}
}

\begin{abstract}
Modern recommender systems heavily rely on high-quality representations learned from high-dimensional sparse data. While significant efforts have been invested in designing powerful algorithms for extracting user preferences, the factors contributing to good representations have remained relatively unexplored. In this work, we shed light on an issue in the existing pair-wise learning paradigm (\textit{i.e.,} embedding collapse), that the representations tend to span a subspace of the whole embedding space, leading to a suboptimal solution and reducing the model capacity. Specifically, we show that alignment of positive pairs is equivalent to a low pass filter causing users/items to collapse to a constant vector. While negative sampling can partially mitigate this issue by acting as a high pass filter to balance the spectrum, leading to an incomplete collapse. \par
To tackle this issue, we present a novel learning paradigm DirectSpec, which directly optimizes the spectrum distribution to ensure that users/items effectively span the entire embedding space. We demonstrate that many self-supervised learning algorithms without explicit negative sampling can be considered as special cases of DirectSpec. Furthermore, we show that optimizing the spectrum inappropriately could also be poisonous to data representation, where the key lies in a dynamic balance between alignment of positive pairs and spectrum balancing. Finally, we propose an enhanced and practical implementation DirectSpec$^+$ to balance the embedding spectrum more adaptively and effectively. We implement DirectSpec$^+$ on two popular recommender models: MF and LightGCN. Our experimental results demonstrate its effectiveness and efficiency over competitive baselines.
\end{abstract}

\ccsdesc[500]{Information systems~Recommender systems}

\keywords{Recommender System, Collaborative Filtering, Embedding Collapse, Embedding Spectrum}

\maketitle

\section{Introduction}
Recommender systems have penetrated into our daily life, we can see them everywhere such as e-commerce \cite{wang2018billion}, short-video \cite{liu2019user}, social network \cite{jiang2012social}, and so on. Collaborative filtering (CF), a fundamental technique for recommendation to discover user preference based on the historical data, has attracted much attention in the last decades. The most extensively used CF technique, matrix factorization (MF) \cite{koren2009matrix} which represents users and items as low dimensional latent vectors and estimates the rating as the inner product between latent vectors, has been the cornerstone of modern recommender systems. Since MF estimates the rating with a simple linear function, subsequent works mostly focus on designing more powerful and complex algorithms to model non-linear user-item relations, including but not limited to multi-layer perceptron (MLP)~\cite{he2017neural}, attention mechanism \cite{kang2018self}, reinforcement learning \cite{zheng2018drn}, transformer \cite{sun2019bert4rec}, diffusion model \cite{wang2023diffusion}, graph neural network (GNN) \cite{ying2018graph,peng2022svd}, etc., and have shown tremendous success.\par 
 
\begin{figure}
\begin{center}
\includegraphics[width=0.75\textwidth]{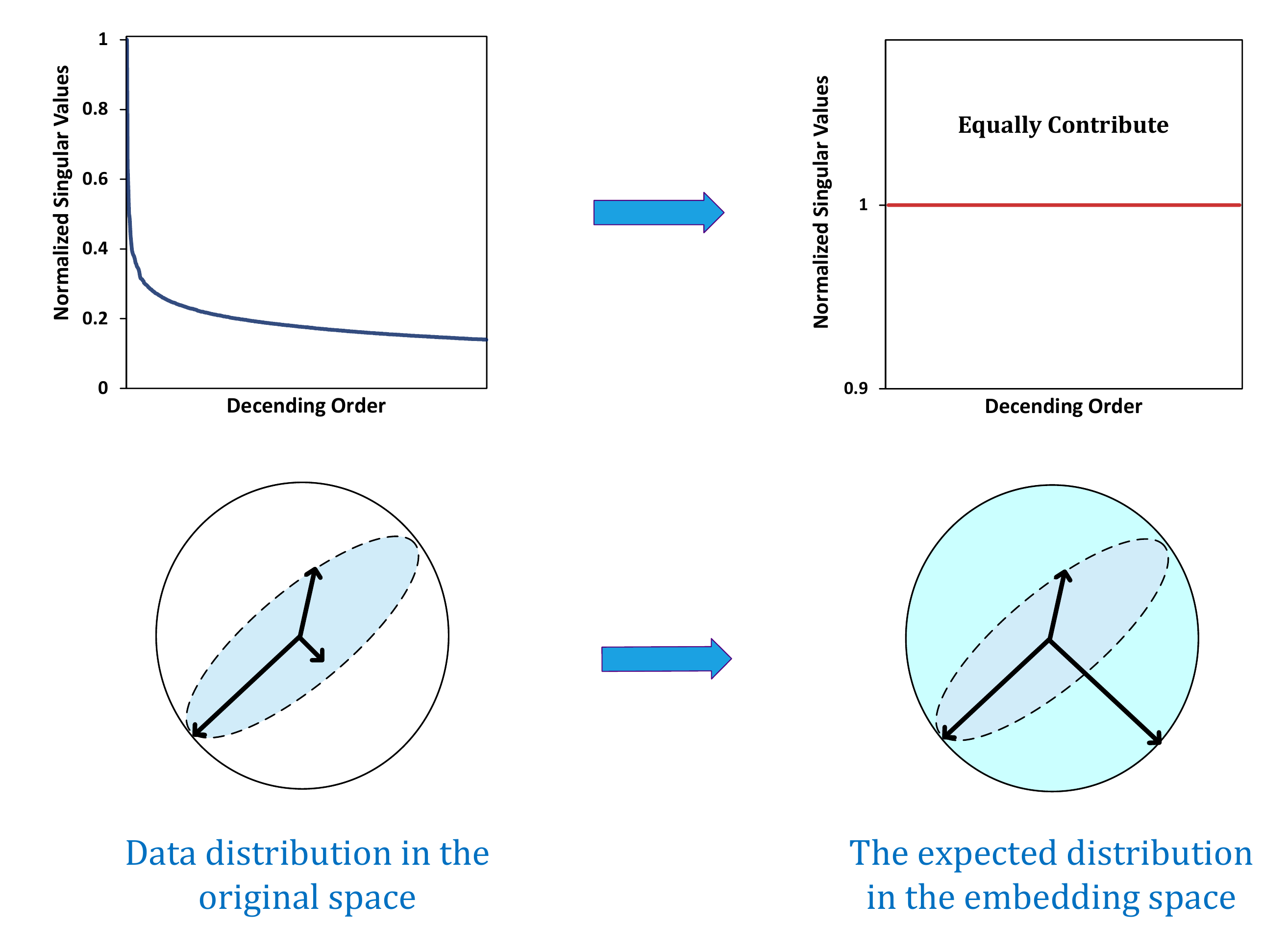}
\caption{An illustration of the data distribution in the original space and expected distribution in the learning embedding space.}
\label{motivation}
\end{center}
\end{figure}

Although different kinds of methods have been proposed, most of them can be considered as MF variants whose goal is to learn low dimensional representations (with dimension $d$) from the high dimensional sparse interaction matrix (with dimension $D\gg d$). Figure \ref{motivation} illustrates the top 500 normalized singular value distribution of the interaction matrix of CiteULike (see Section 5.1 for data description). We observe that users/items are predominantly distributed along only a few dimensions in the original space while most dimensions barely contribute (\textit{i.e.,} with singular values close to 0) to the representations. Thus, when users and items are mapped into a more compact embedding space, it is expected that redundant dimensions are all removed and each dimension contributes to the user/item representations as equally and uniformly as possible (\textit{i.e.,} the representations make full use of the embedding space). \par

Existing recommendation methods mostly utilize positive (observed) and negative (unobserved) samples for optimization. Here, a simple yet fundamental question arises: Can user/item representations of existing works make full use of all dimensions? Unfortunately, by analyzing the spectrum of the embedding matrix, we empirically and theoretically show that users/items tend to span a subspace of the whole embedding space (with dimension $d'<d$), where the embeddings collapse along all (complete collapse) or certain dimensions (incomplete collapse). Particularly, alignment of positive pairs is equivalent to a low pass filter, where the representations tend to collapse to a constant vector. Negative sampling is commonly used to optimize recommendation algorithms without causing an explicit embedding collapse by pushing away the unobserved user-item pairs, and we show that it is equivalent to a high pass filter decelerating the speed of collapse by balancing the embedding spectrum. However, the effectiveness of negative sampling is only limited to certain loss functions such as Bayesian personalized ranking (BPR) \cite{rendle2009bpr} and binary cross-entropy (BCE) loss~\cite{he2017neural}, where collapse over certain dimensions still exists.\par

In this work, we tackle the embedding collapse issue from a spectral perspective. We observe that the extent of the collapse is closely related to the spectrum distribution of the embedding matrix. Specifically, only a single singular value dominates when the representation completely collapses, whereas the singular values are uniformly distributed when the representations make full use of the embedding space. Inspired by this observation, we propose to directly optimize the spectrum distribution (dubbed DirectSpec) ensuring that all dimensions equally contribute to the representations. We prove that DirectSpec can completely prevent the embedding collapse and provide a practical implementation with a complexity as $\mathcal{O}(B^2d)$ where $B$ is the batch size. In addition, we demonstrate that advanced self-supervised learning (SSL) algorithms such as Barlow Twins, LogDet, and spectral contrastive loss without explicit negative sampling can be considered as special cases of DirectSpec. Furthermore, we show that optimizing the spectrum distribution could also be poisonous to data representation, where the key lies in a dynamic balance between alignment of positive pairs and spectrum balance. Based on this observation, we point out the limitations of DirectSpec and propose an enhanced variant DirectSpec$^+$ to balance the embedding spectrum more adaptively and effectively. Finally, we implement DirectSpec and DirectSpec$^+$ on two popular baselines: MF \cite{koren2009matrix} and LightGCN \cite{he2020lightgcn}, and experimental results show that DirectSpec$^+$ improves BPR and LightGCN by up to 52.6\% and 42.4\% in terms of nDCG@10, respectively. The contributions of this work can be summarized as follows:
\begin{itemize}[leftmargin=10pt]
\item We theoretically and empirically show that existing recommendation methods suffer from embedding collapse, where the representations tend to fall into a subspace of the whole embedding space, and analyze the mechanisms causing this issue. 

\item We propose a novel learning paradigm DirectSpec which directly optimizes the spectrum distribution, ensuring that users/items effectively span the entire embedding space. We empirically and theoretically show that DirectSpec can prevent embedding collapse.

\item We demonstrate that SSL methods such as Barlow Twins, LogDet, and spectral contrastive loss without explicit negative sampling can be considered as special cases of DirectSpec.
  
\item Extensive experimental results on three datasets demonstrate the efficiency and effectiveness of our proposed methods. 
\end{itemize}  

\section{Preliminaries}
\subsection{Pairwise Learning for CF}
Given the user set $\mathcal{U}$ and item set $\mathcal{I}$, the interaction matrix is defined as $\mathbf{R}\in \{0, 1\}^{\left | \mathcal{U} \right | \times \left | \mathcal{I} \right |}$, the observed interactions are represented as $\mathbf{R}^+=\{r_{ui}=1|u\in\mathcal{U}, i\in\mathcal{I}\}$. Users and items are first mapped to low dimensional vectors $\mathbf{H}\in\mathbb{R}^{(\left | \mathcal{U} \right | + \left | \mathcal{I}\right|)\times d}$ through an encoder, where the encoder can be as simple as a linear mapping \cite{koren2009matrix} or advanced algorithms such as MLPs \cite{he2017neural}, GNNs \cite{wang2019neural}, etc. Let $\mathbf{h}_u$ and $\mathbf{h}_i$ be the $u$'s and $i$'s representations, respectively. The goal of CF is to predict unobserved interactions $\mathbf{R}^-=\{r_{ui}=0|u\in\mathcal{U}, i\in\mathcal{I}\}$  estimated as the inner product between the user and item representations: $\hat{r}_{ui}=\mathbf{h}_u^T \mathbf{h}_i$. The model parameters $\Theta$ are optimized through a loss function $\mathcal{L}$ formulated as follows: 
\begin{equation}
{\rm arg} \mathop{{\rm min}}\limits_{\mathbf{\Theta}} \mathcal{L}(\hat{r}_{ui},r_{ui}). 
\label{loss}
\end{equation} 
The loss function measures the difference between the estimated score and the ground truth. For instance, BPR and BCE loss are extensively used for CF methods:
\begin{equation}
\begin{aligned}
&\mathcal{L}_{BPR}=\sum_{(u,i)\in{\mathbf{R}}^+,(u,j)\in{\mathbf{R}}^-}-\ln \sigma(\hat{r}_{ui}-\hat{r}_{uj}),\\
&\mathcal{L}_{BCE}= \sum_{u\in\mathcal{U},i\in \mathcal{I}}- r_{ui}\ln \sigma(\hat{r}_{ui})- \left( 1-r_{ui}\right)\ln \left( 1-\sigma(\hat{r}_{ui})\right),
\end{aligned}
\label{bpr_entropy}
\end{equation}
where $\sigma(\cdot)$ is the sigmoid function. BPR loss maximizes the difference between observed and unobserved user-item pairs, while BCE loss directly pulls the observed pairs close and pushes the unobserved ones away from each other. The embeddings are a low dimensional approximation of the sparse high dimensional interaction matrix, thus they should contain diverse and rich information representing the user-item relations. Rank is a commonly used metric to measure the dimension of a matrix, while it fails to tell the difference between the embedding matrix (\romannum{1}): with a `sphere' distribution that users/items are uniformly distributed in each dimension of the space and (\romannum{2}): with a `spheroid' distribution that users/items are predominantly distributed over some dimensions and insignificantly distributed over other dimensions. Although both matrices have the same rank, apparently (\romannum{1})  contains more diverse information than (\romannum{2}). Here, we introduce another tool:
\begin{definition}
\textbf{Effective Rank.} Given singular values of the embedding matrix $\mathbf{H}$: $\sigma_1\geq\sigma_2\geq\cdots\geq\sigma_d\geq0$, let $p_k=\frac{\sigma_k}{\sum_k \sigma_k}$, then the effective rank is defined as follows:
\begin{equation}
erank(\mathbf{H})=exp\left( H\left(p_1,\cdots, p_d \right) \right),
\label{erank}
\end{equation}
where $H\left(p_1,\cdots, p_d \right)=-\sum_k p_k \log p_k$ is the Shannon entropy. 
\end{definition}
Compared with rank, erank takes the singular value distribution into consideration: the more uniform (sharper) of the distribution, the higher (lower) of the erank \cite{roy2007effective}. The embedding matrix contains the least information when there is only one leading non-zero singular values ($\sigma_2=\cdots=\sigma_d=0$, and $erank(\mathbf{H})=rank(\mathbf{H})=1$), while erank is maximized when each dimension equally contributes to the representations: $\sigma_1=\cdots=\sigma_d$ ($erank(\mathbf{H})=rank(\mathbf{H})=d$). 

\subsection{Graph Representation and Graph Filtering}
The interactions can be represented as a graph. Consider a graph $\mathcal{G}=(\mathcal{V},\mathcal{E},\mathbf{A})$, where the node set contains all users and items: $\mathcal{V}=\mathcal{U}\cup\mathcal{I}$, the edge set is represented by observed interactions: $\mathcal{E}=\mathbf{R}^+$, $\mathbf{A}$ is an adjacency matrix. Suppose $\mathbf{A}$ is normalized with the eigenvalue $|\lambda_k|\leq1$, then: 
\begin{definition}
\textbf{Graph Filtering.} Let $1=\lambda_1>\lambda_2>\cdots>\lambda_n=-1$ be the eigenvalues of $\mathbf{A}$ where $n=|\mathcal{V}|$. The components closer to $\lambda_1$ and $\lambda_n$ correspond to a lower and higher frequency, respectively. $\mathcal{F}^L(\mathbf{A})$ is a low pass filter if $|\mathcal{F}^L(\lambda_i)|>|\mathcal{F}^L(\lambda_j)|$ for $\lambda_i>\lambda_j$ and $\mathcal{F}^H(\mathbf{A})$ is a high pass filter if $|\mathcal{F}^H(\lambda_i)|>|\mathcal{F}^H(\lambda_j)|$ for $\lambda_i<\lambda_j$.  
\end{definition}
Note that $\mathcal{F}^L(\mathbf{A})$ is defined as a low pass filter as it emphasizes more on the lower frequencies and similarly for the definition of the high pass filter. Please refer to \cite{sandryhaila2014discrete} for detailed introduction of graph frequency.

\begin{figure} \centering 
\subfigure[] {  
\includegraphics[width=0.31\columnwidth]{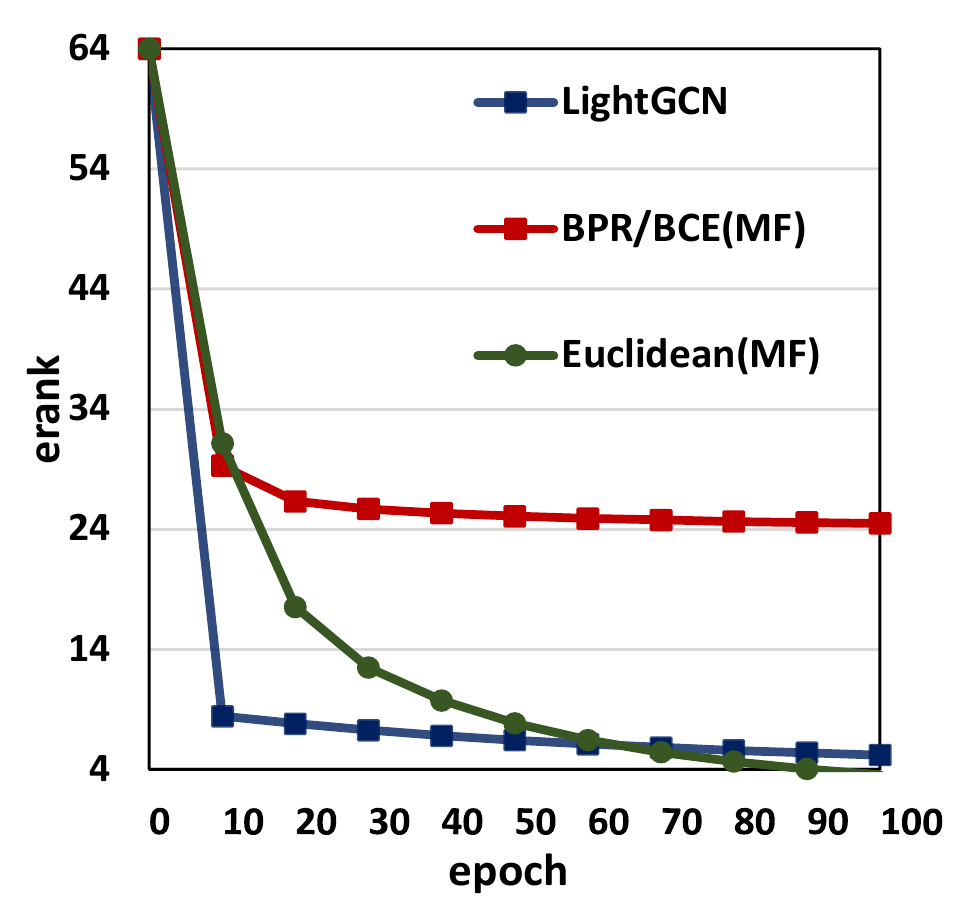} 
}
\subfigure[] {  
\includegraphics[width=0.31\columnwidth]{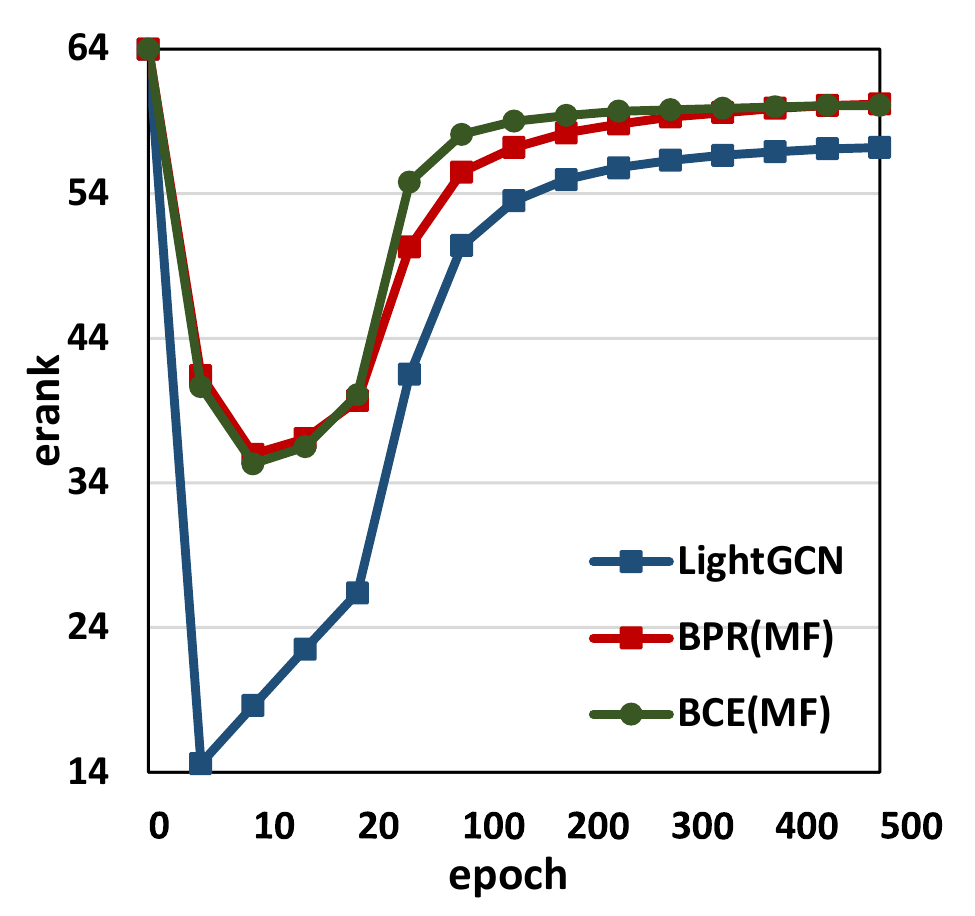} 
}
\subfigure[] {  
\includegraphics[width=0.31\columnwidth]{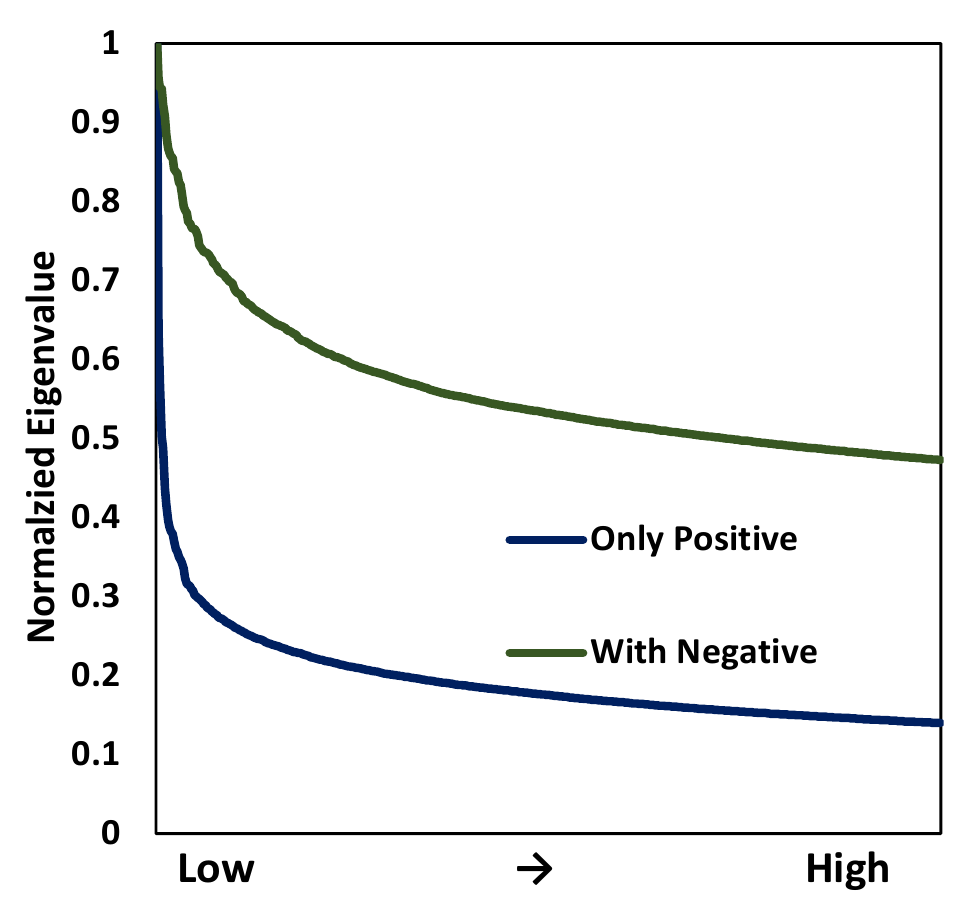} 
}

\caption{(a) complete collapse, (b) incomplete collapse, (c) how negative sampling can balance the spectrum on Yelp: the normalized eigenvalue distribution (Top 500) of $\mathbf{A}-\mathbf{\bar{A}}$.}           
\label{collapse}
\end{figure}

\section{Embedding Collapse in CF}
\subsection{Complete Collapse}
Let us first consider a generic loss function only optimizing alignment of positive pairs (\textit{i.e.,} minimizing their distance): $\mathcal{L}(\hat{r}_{ui},r_{ui}), (u,i)\in\mathbf{R}^+$.
\begin{proposition} 
Suppose $G$ is connected, then $\mathbf{h}_k\approx \mathbf{h}_z$ for arbitrary nodes $k$ and $z$ when $\mathcal{L}$ completely converges. 
\end{proposition}
By calculating the gradient over the embeddings, the parameters are updated through stochastic gradient descent (SGD) as follows:
\begin{equation}
\begin{aligned}
&\mathbf{H}^{(l+1)} =\mathbf{H}^{(l)} - \alpha\frac{\partial \mathcal{L}}{\partial \mathbf{H}^{(l)}}=\mathbf{H}^{(l)} + \alpha \mathbf{A} \mathbf{H}^{(l)},\\
&\mathbf{h}^{(l+1)}_u =\mathbf{h}^{(l)}_u - \alpha\frac{\partial \mathcal{L}}{\partial \mathbf{h}_u^{(l)}}=\mathbf{h}^{(l)}_u +\alpha \sum_{(u,i)\in\mathbf{R}^+} \mathbf{A}_{ui}  \mathbf{h}^{(l)}_i,
\end{aligned}
\label{log_pos_update}
\end{equation}
where $\alpha\in (0,1)$, $\mathbf{A}_{ui}=-\frac{\partial \mathcal{L}}{\partial \hat{r}_{ui} }$ for $(u,i)\in\mathbf{R}^+$. Ideally, if $\frac{\partial \mathcal{L}}{\partial \hat{r}_{ui} }$ remain unchanged during training such as the Euclidean distance loss: $\mathcal{L}=\sum \left \|  \mathbf{h}_u - \mathbf{h}_i\right \|^2_2$, then Equation (\ref{log_pos_update}) can be rewritten as:
\begin{equation}
\mathbf{H}^{(l)}=\left( \mathbf{I} + \alpha \mathbf{A}\right)\mathbf{H}^{(l-1)}=\cdots =\left( \mathbf{I} + \alpha \mathbf{A}\right)^l\mathbf{H}^{(0)}.
\label{converge_collapse}
\end{equation}
Suppose $\mathbf{A}$ is normalized such as $|\lambda_k|\leq 1$, then  $(\mathbf{I}+\alpha \mathbf{A})^l$ is equivalent to a low pass filter, as the eigenvalue of $\mathbf{I}+\alpha \mathbf{A}$ is $1+\alpha\lambda_k \in [1-\alpha,1+\alpha]$ with $1-\alpha>0$ and $x^l$ is monotonically increasing for $l>0$ and $x>0$. According to the spectral decomposition with $\mathbf{v}_k$ as the eigenvector:
\begin{equation}
\left( \mathbf{I} + \alpha \mathbf{A}\right)^l=\sum_k (1+\alpha\lambda_k)^l\mathbf{v_k}\mathbf{v_k}^T,
\end{equation}
it is obvious that $\frac{(1+\alpha\lambda_k)^l}{(1+\alpha\lambda_1)^l}\rightarrow0$ ($k\neq 1$) as $l$ is large enough, resulting in rank$((\mathbf{I} + \alpha \mathbf{A})^l)\rightarrow1$, then: 
\begin{equation}
{\rm rank} \left( \mathbf{H}^{(l)} \right)\leq {\rm min}\left( {\rm rank}\left( (\mathbf{I} + \alpha \mathbf{A})^l \right),{\rm rank}\left(\mathbf{H}^{(0)}\right) \right)= 1,
\end{equation}
showing that all nodes have the same embedding representations. In reality, all users/items would not completely converge to a constant vector as: (\romannum{1}): $\mathcal{G}$ is not connected and (\romannum{2}): the magnitude of the gradient $\mathbf{A}_{ui}$ changes during training. For instance, $\mathbf{A}_{ui}=1-\sigma(\hat{r}_{ui})$ for the log loss: $\mathcal{L}=-\sum\ln\sigma(\hat{r}_{ui})$ where $\mathbf{A}_{ui}$ tends to vanish as training leads to $\sigma(\hat{r}_{ui})\rightarrow 1$, while $\mathbf{A}_{ui}=1$ for the Euclidean distance loss. Thus, the embeddings tend to collapse more quickly on the Euclidean distance loss than the log loss. In Figure \ref{collapse} (a), we evaluate LightGCN and MF on how the embeddings completely collapse on Yelp when only optimizing alignment of positive pairs, where the parameters are initialized with Xavier initialization and $d=64$. We have the following observations:
 
\begin{itemize}[leftmargin=10pt]

\item Before the training starts, the embeddings are uniformly distributed in the embedding space since $erank(\mathbf{H})\approx d$.

\item The erank monotonically declines on both models, indicating that the embeddings collapse as the training starts and the issue becomes more serious as training proceeds.

\item The erank decreases more rapidly on LightGCN than MF. Consider a $K$-layer LightGCN, then Equation (\ref{converge_collapse}) can be rewritten as:
\begin{equation}
\mathbf{H}^{(l)} =\left( \mathbf{I} + \alpha \mathbf{A}\right)^l\sum_{k=0}^K \mathbf{A}^k \mathbf{E}.
\end{equation}   
Here, $\mathbf{E}$ is the initial stacked user/item embeddings sent to the encoder; we ignore the difference between the adjacency matrix used in Equation (\ref{converge_collapse}) and LightGCN for simplicity. We can see that increasing the layer $K$ causes the loss function to converge faster, indicating that the over-smoothing in GNN \cite{li2018deeper} aggravates the collapse issue. 

\item The erank tends to converge to a value larger than 1 due to the two reasons we mentioned above. By comparing the Euclidean distance loss (MF) and log loss (MF), we can see the erank of the Euclidean distance loss with stable gradients drops more quickly than that of log loss, which further verifies our analysis. 
 
\end{itemize}

\subsection{Incomplete Collapse}
Existing recommendation algorithms mostly exploit negative sampling for optimization. Naturally, we raise a question: Can existing pair-wise learning paradigm completely prevent the collapse issue? Similarly, we evaluate LightGCN (with BPR), MF (with BPR), and MF (with BCE) on Yelp, and show how erank changes as training proceeds in Figure \ref{collapse} (b). The erank plunges significantly in the first place, showing a trend similar to Figure \ref{collapse} (a). Gradually, the erank increases and tends to converge to a value lower than $d$. Overall, despite its partial effectiveness, (\romannum{1}): the fluctuation of erank, especially the significant drop at the start of training and (\romannum{2}): its convergence to a value significantly lower than $d$, demonstrate that negative sampling cannot fully prevent the collapse issue. We attempt to analyze the observations in Figure \ref{collapse} (b). Consider a generic loss function optimizing both positive and negative samples : $\mathcal{L}(\hat{r}_{ui},r_{ui}), (u,i)\in\mathbf{R}$. The updating rule via SGD can be formulated as follows:

\begin{equation}
\begin{aligned}
&\mathbf{H}^{(l+1)}=\mathbf{H}^{(l)} + \alpha \left( \mathbf{A}- \mathbf{\bar{A}} \right)\mathbf{H}^{(l)},\\
&\mathbf{h}^{(l+1)}_u =\mathbf{h}^{(l)}_u + \alpha \sum_{(u,i)\in\mathbf{R}^+} \mathbf{A}_{ui}  \mathbf{h}^{(l)}_i-\alpha \sum_{(u,j)\in\mathbf{R}^-} \mathbf{\bar{A}}_{uj}  \mathbf{h}^{(l)}_j,
\end{aligned}
\label{bce_update}
\end{equation}
where $\mathbf{\bar{A}}_{uj}=\frac{\partial \mathcal{L}}{\partial \hat{r}_{uj} }$ for $(u,j)\in\mathbf{R}^{-}$. Note that the gradients for positive and negative samples always have the opposite direction. Here, $ \mathbf{I} + \alpha \left( \mathbf{A}-\mathbf{\bar{A}} \right)$ can be disentangled to a low pass $\mathbf{I}+\alpha\mathbf{A}$ and a high pass filter $\mathbf{I}-\alpha\mathbf{\bar{A}}$, where a high pass filter can balance the embeddings spectrum by reducing the weights on low frequency and raising the importance of high frequency components, leading to a more uniform distribution.  We visualize the spectrum of $\mathbf{A}-\mathbf{\bar{A}}$ in Figure \ref{collapse} (c), where we fix the elements of $\mathbf{A}$ and $\mathbf{\bar{A}}$ to 1 and normalize them separately. $\mathbf{\bar{A}}=0$ (zero matrix) for `Only positive', and $\mathbf{\bar{A}}$ contains all unobserved interactions for `With negative'. It is obvious that $\mathbf{A}-\mathbf{\bar{A}}$ has a more balanced spectrum than $\mathbf{A}$, leading to a higher erank.

\begin{proposition} 
Negative sampling alone leads to collapse. 
\end{proposition}

Suppose we only focus on the negative sampling with the updating rule as: $\mathbf{H}^{(l+1)}=\mathbf{H}^{(l)} - \alpha \mathbf{\bar{A}} \mathbf{H}^{(l)}$. Ideally (\textit{i.e.,} with the gradients remaining unchanged), the updating rule can be rewritten as follows:
\begin{equation}
\mathbf{H}^{(l)} =\left( \mathbf{I} - \alpha \mathbf{\bar{A}}\right)^l\mathbf{H}^{(0)}.
\label{negative_collapse}
\end{equation}
Similar to the analysis in Section 3.1, the high frequency components gradually dominate ($\frac{(1-\alpha\lambda_k)^l}{(1-\alpha\lambda_n)^l}\rightarrow0$) as $l$ becomes larger, and eventually ${\rm rank} \left( \mathbf{H}^{(l)} \right)\rightarrow 1$. Then, another question arises: If both positive and negative sampling lead to collapse, how do existing methods avoid the collapse?\par

We further zoom in on the gradient over BCE loss. The gradient over the positive sample $\mathbf{A}_{ui}$: $1-\sigma(\hat{r}_{ui})$ is close to 1 at the beginning of training, causing the embeddings to collapse. Since positive pairs are more likely to be sampled due to the data sparsity, the gradients vanish faster as they are pulled close: $\mathbf{A} \rightarrow 0$, allowing the high pass filter to balance the spectrum. As the negative samples are sampled and pushed away, we have: $\mathbf{\bar{A}}_{uj}=\sigma(\hat{r}_{uj}) \rightarrow0$, making the representations converge: $\mathbf{H}^{(l+1)} =\mathbf{H}^{(l)}$ and leading to an incomplete collapse. For BPR loss, since the magnitude of gradients is $1-\sigma(\hat{r}_{ui}-\hat{r}_{uj})$ for both positive and negative pairs, positive pairs being optimized causes the gradients for negative pairs to vanish as well, making the high pass filter have a weaker effect balancing the spectrum and also explaining why BPR(MF) converges more slowly than BCE(MF). To conclude, negative sampling is not designed to prevent the collapse issue, gradient vanishing on log loss is the `X factor' to alleviate the complete collapse, and is also the reason causing the incomplete collapse as well. Therefore, a more effective design proved to completely solve the collapse issue is required.

\section{Methodology}
\subsection{Directly Balancing Embedding Spectrum (DirectSpec)}
The spectrum distribution can directly reflect the extent of embedding collapse. Intuitively, a uniform distribution indicates that different dimensions equally contribute to the embeddings and leads to a high effective rank. To this end, we consider the following Lagrangian:  
\begin{equation}
\mathcal{L}=f\left( \sigma_1,\cdots,\sigma_d  \right)+ \beta \left( \sqrt{\sum_i \sigma_i^2}-c  \right),
\label{balance_spec}
\end{equation}
with the minimum reached at $\sigma_1=\cdots=\sigma_d=\frac{c}{\sqrt{d}}$, where the singular value is bounded by the $L_2$ ball with radius $c$.

\begin{proposition} 
Let $\frac{\partial f}{\partial\sigma_i}=g_i(\cdot)$. The minimum of Equation (\ref{balance_spec}) is achieved at $\sigma_1=\cdots=\sigma_d=\frac{c}{\sqrt{d}}$ as long as $g_i(\sigma_j)=g_j(\sigma_j)$ for $i,j\in\{1,\cdots,d \}$.
\end{proposition}
By calculating the derivative over $\sigma_i$, we get:
\begin{equation}
\left\{
             \begin{array}{lr}
             g_1(\sigma_1)+ 2\beta\sigma_1=0    \\\\
             \qquad \;\; \cdots                             \\\\
             g_d(\sigma_d)+ 2\beta\sigma_d=0  \\\\
             
             s.t. \sum_i \sigma_i^2=c^2. \\ 
             \end{array}
\right.
\end{equation}
Obviously, $\sigma_1=\cdots=\sigma_d$ as long as $g_i(\cdot)$ has the same form for all singular values. For instance, a simple design is $f\left( \sigma_1,\cdots,\sigma_d  \right)=\sum_i\sigma_i$ where $\beta=-\frac{\sqrt{d}}{2c}$.\par

While the algorithm above has a provable guarantee to prevent the collapse issue, the efficiency is limited as it requires computing the singular values with the complexity as: $\mathcal{O}(\left | \mathcal{U} \right |d^2 + \left | \mathcal{I}\right| d^2)$.

\begin{figure} \centering 
\subfigure[] {  
\includegraphics[width=0.45\columnwidth]{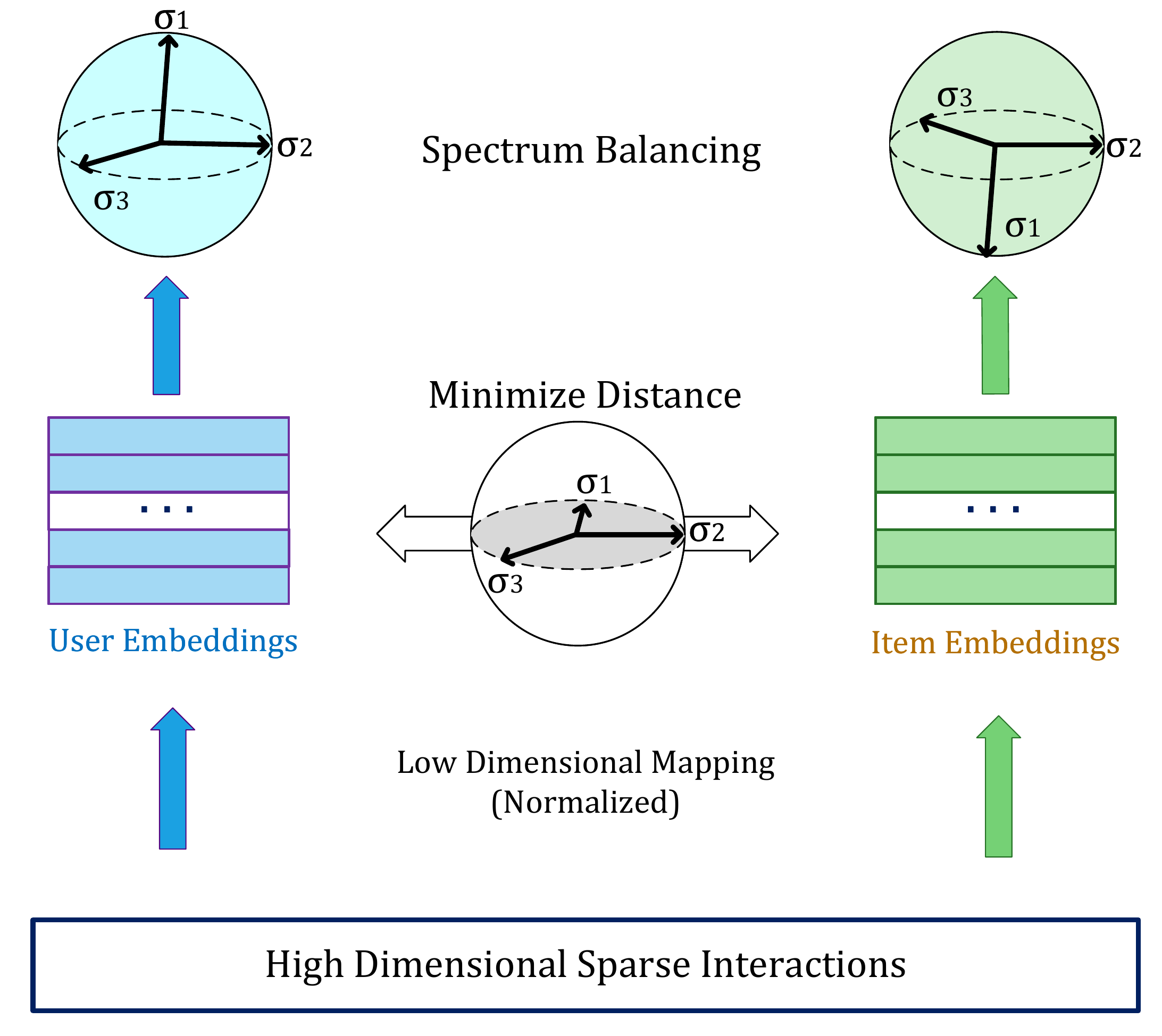} 
}
\subfigure[] {  
\includegraphics[width=0.45\columnwidth]{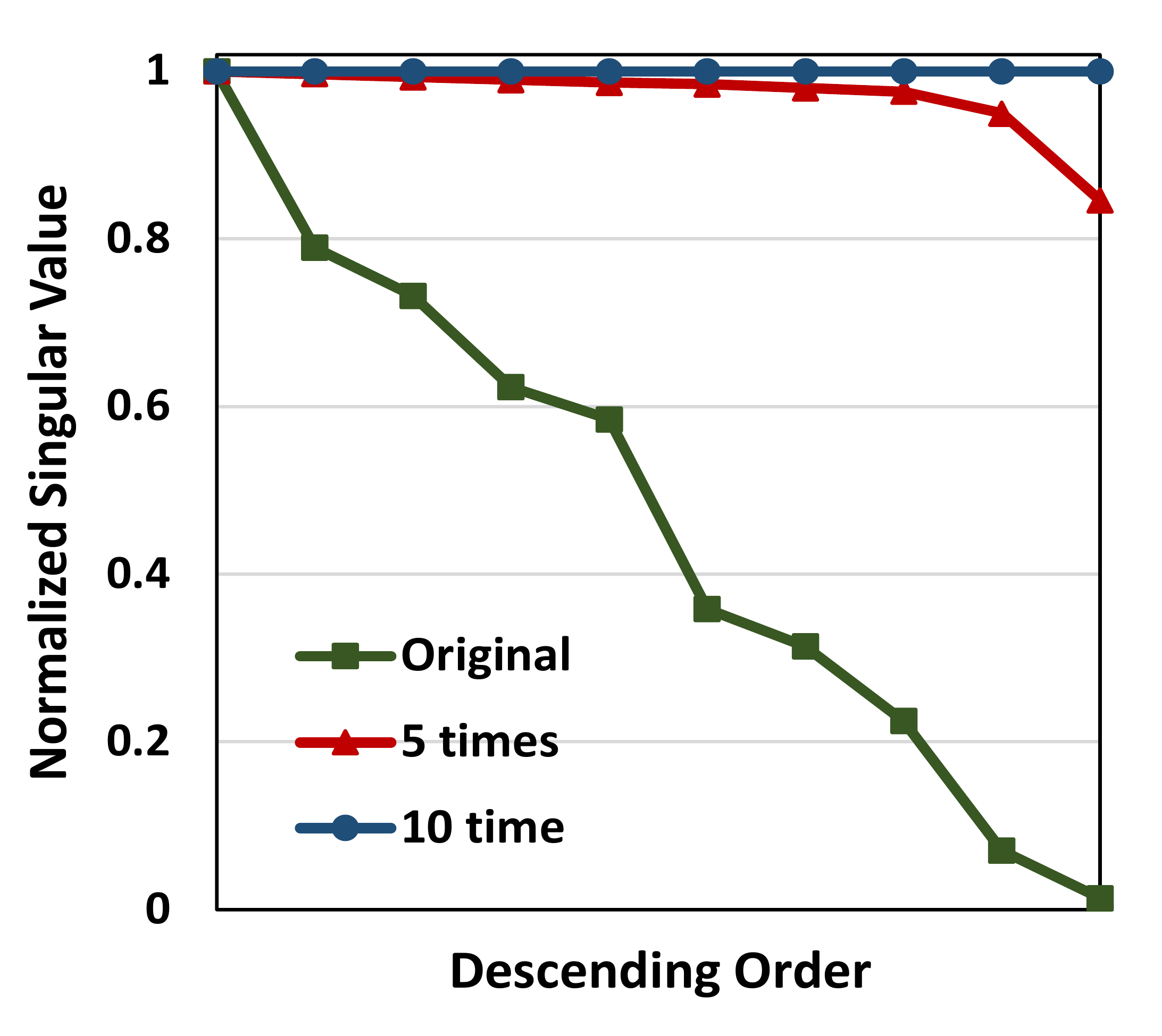} 
}
\caption{(a): An illustration of the proposed DirectSpec, (b): A toy example running DirectSpec on a randomly generated matrix with torch.randn(10,10) $\{5, 10\}$ times.  }           
\label{model}
\end{figure}

\begin{proposition}
When $f\left( \sigma_1,\cdots,\sigma_d  \right)=\sum_i\sigma_i^{2K}$, $K\in\mathbb{N}_{+}$ and $K\geq2$, we can balance the spectrum with the following Equation:
\begin{equation}
\mathbf{H}\leftarrow \mathbf{H}- \alpha \left(\mathbf{H}\mathbf{H}^T\right)^{K-1}\mathbf{H}.
\label{proposition_3_update}
\end{equation}
\end{proposition}
\begin{algorithm}
\SetAlgoNoLine
\SetKwInOut{Input}{Input}
\caption{DirectSpec}
\Input{Embedding matrix $\mathbf{H}$; batch size $B$; hyperparameter $\alpha\in\mathbb{R}^+$}
sample the users and items from $\mathbf{R}^+$: $\mathcal{U}_B$, $\mathcal{I}_B$ \;
generate the normalized embeddings: $\mathbf{H}_{B}^U$, $\mathbf{H}_{B}^I$\;
$\mathbf{H}_{B}^U\leftarrow\mathbf{H}_{B}^U-\alpha  \mathbf{H}_{B}^U{\mathbf{H}_{B}^U}^T \mathbf{H}_{B}^U$\;
$\mathbf{H}_{B}^I\leftarrow\mathbf{H}_{B}^I-\alpha  \mathbf{H}_{B}^I{\mathbf{H}_{B}^I}^T \mathbf{H}_{B}^I$\;
update parameters via $\mathcal{L}(\hat{r}_{ui},r_{ui}), (u,i)\in\mathbf{R}^+$
\label{directspec}
\end{algorithm}
According to SVD: $\mathbf{H}=\mathbf{P}diag(\sigma_i)\mathbf{Q}^T$ with $\mathbf{P}$, $\mathbf{Q}$, and $diag(\cdot)$ as the left, right stacked singular vectors, and the diagonalization operator, respectively. It is obvious that:
\begin{equation}
\left( \mathbf{H} \mathbf{H}^T \right)^{K}=\left(\mathbf{P}diag\left(\sigma_i^2 \right)\mathbf{P}^T \right)^{K}=\mathbf{P}diag\left(\sigma_i^{2K} \right)\mathbf{P}^T ,
\end{equation}
due to the orthogonality of singular vectors: $\mathbf{Q}^T\mathbf{Q}=\mathbf{I}$, $\mathbf{P}^T\mathbf{P}=\mathbf{I}$. Then, we can make the following observation:
\begin{equation}
f\left( \sigma_1,\cdots,\sigma_d  \right)=\sum_i\sigma_i^{2K}=\Tr \left( \mathbf{P}diag\left(\sigma_i^{2K} \right)\mathbf{P}^T \right)=\Tr \left( \left( \mathbf{H} \mathbf{H}^T \right)^{K} \right),
\end{equation}
where the trace $\Tr(\cdot)$ is equal to the sum of eigenvalues. Then, the embedding can be updated via SGD as follows:
\begin{equation}
\mathbf{H}\leftarrow \mathbf{H} - \alpha \frac{\partial \Tr \left( \left( \mathbf{H} \mathbf{H}^T \right)^{K} \right)}{\partial \mathbf{H}}=\mathbf{H} - \alpha K\left(\mathbf{H}\mathbf{H}^T\right)^{K-1}\mathbf{H}.
\end{equation}
By simply setting $\alpha\leftarrow \alpha K$, we obtain Equation (\ref{proposition_3_update}). Note that we can ignore the constraint term (\textit{i.e.,} $\sum_i \sigma_i^2=c^2$) as it can be replaced by the normalization: $\mathbf{H}_i\leftarrow \frac{\mathbf{H}_i}{\left|\mathbf{H}_i\right|}$ where $\mathbf{H}_i$ represents an arbitrary row of $\mathbf{H}$, simply due to the following relation:

\begin{equation}
\sum_i \sigma_i^2= \Tr \left( \mathbf{P}diag\left(\sigma_i^{2} \right)\mathbf{P}^T \right)=\Tr(\mathbf{H}\mathbf{H}^T)=\sum_i \mathbf{H}_i \mathbf{H}_i^T=d,
\label{norm_stable}
\end{equation}
where the diagonal element $ \mathbf{H}_i \mathbf{H}_i^T=1$. To further shed light on how Equation (\ref{proposition_3_update}) balances the spectrum, we can rewrite it as follows:
\begin{equation}
\begin{aligned}
\mathcal{F}\left( \mathbf{H}\right)=\mathbf{H}-\alpha \left( \mathbf{H}\mathbf{H}^T\right)^K \mathbf{H}&=\mathbf{P}diag\left(\sigma_i\right)\mathbf{Q}^T-\alpha \mathbf{P}diag\left(\sigma_i^{2K+1}\right)\mathbf{Q}^T,\\
&=\mathbf{P}diag\left(\sigma_i\left(1-\alpha\sigma^{2K}_i\right) \right)\mathbf{Q}^T,
\end{aligned}
\label{alg1_svd}
\end{equation}
where we redefine $K\geq 1$ for simplicity. As the function $\sigma_i^{2K}$ increases faster than $\sigma_i$, $1-\alpha\sigma^{2K}_i>1-\alpha\sigma^{2K}_j$ for $\sigma_i<\sigma_j$ as long as $\alpha$ is properly set (\textit{i.e.,} $1-\alpha\sigma_1^{2K}>0$). Thus, $1-\alpha\sigma_i^{2K}$ can be considered as rescaled factors: the larger singular values are multiplied by smaller weights, leading to a balanced distribution.  Note that each update would not shrink the singular values as the embedding normalization keeps their $L_2$ norm unchanged as shown in Equation (\ref{norm_stable}). According to the definition of entropy and erank, a more uniform distribution results in a higher erank, indicating that $erank(\mathcal{F}\left( \mathbf{H}\right))\geq erank(\mathbf{H})$ where $=$ holds when $erank(\mathbf{H})=d$. To further enhance the efficiency of DirectSpec, we set $K=1$ and adopt a batch training strategy instead of updating the whole $\mathbf{H}$. We can integrate DirectSpec as a spectrum-normalization layer in the recommendation algorithm as formulated in Algorithm \ref{directspec} and as illustrated in Figure \ref{model} (a), where we do not need to explicitly sample the negative pairs, saving the training cost. To demonstrate and validate the effectiveness of DirectSpec, we run a toy example as shown in Figure \ref{model} (b).


\subsection{Spectrum Balancing Could be Poisonous} 

\begin{figure} \centering 
\subfigure[] {  
\includegraphics[width=0.43\columnwidth]{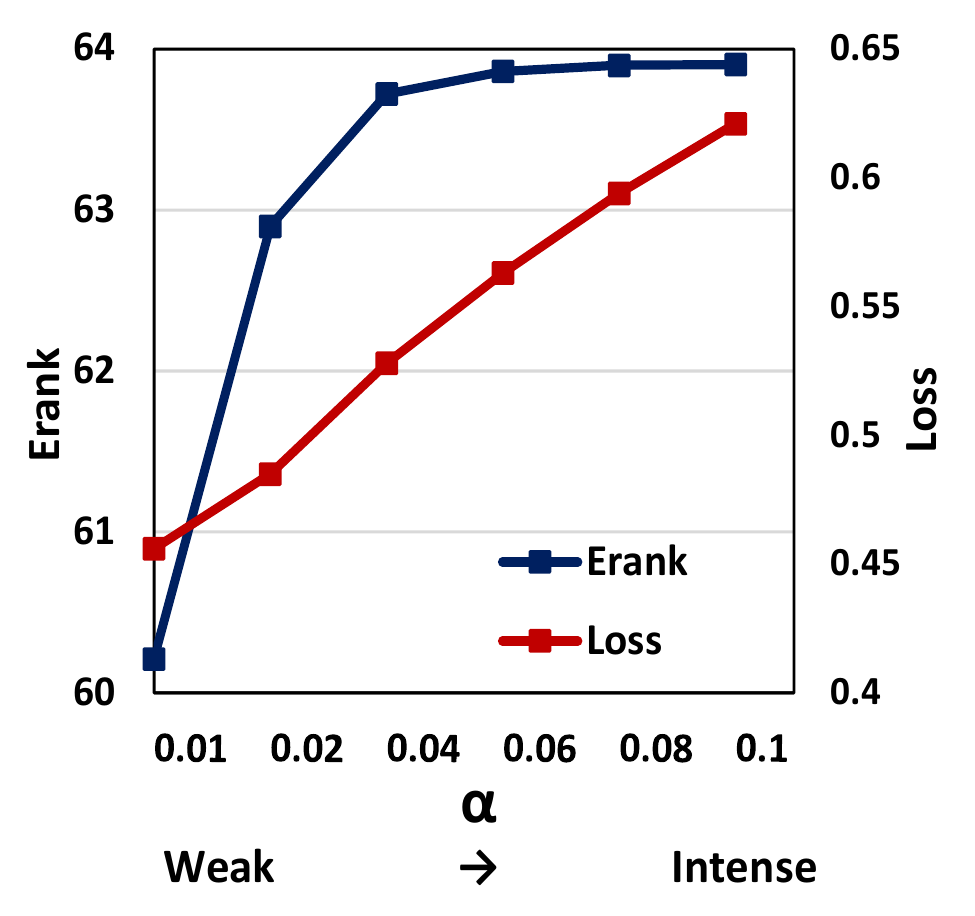} 
}
\subfigure[] {  
\includegraphics[width=0.43\columnwidth]{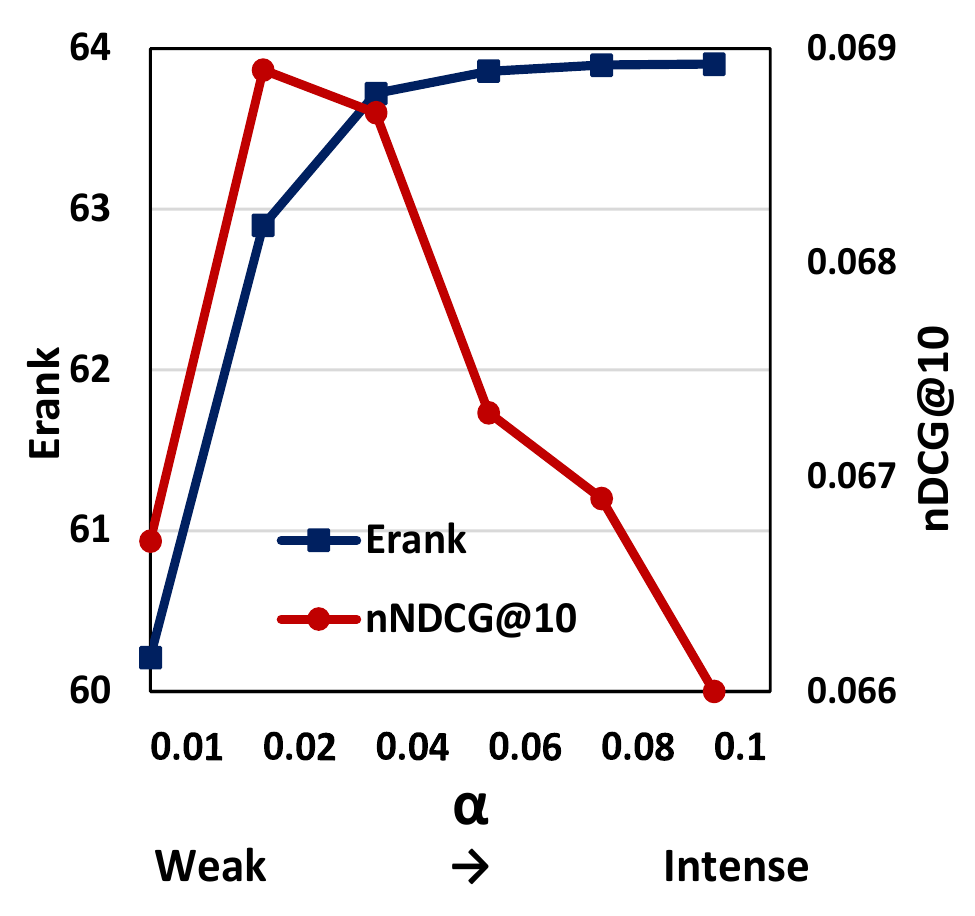} 
}
\caption{How (a): the effective rank and training loss (b): the effective rank and accuracy change with $\alpha$.}           
\label{poison}
\end{figure}

\begin{proposition}
The goal of spectrum balancing and alignment of positive pairs is contradictory.
\end{proposition}
Suppose the user $u$ and $v$ both interacted with item $i$. Let $\mathbf{H}_U$ denote the embedding matrix for users, then $\mathbf{H}_U\mathbf{H}_U^T=\mathbf{I}$ when the perfectly balanced spectrum is achieved: $\sigma_1=\cdots=\sigma_d$, indicating that $u$ and $v$ are orthogonal in the embedding space: $\mathbf{h}_u^T \mathbf{h}_v=0$. On the other hand, $i$ is optimized to be parallel to both $u$ and $v$: $\mathbf{h}_u^T\mathbf{h}_i=1$ and $\mathbf{h}_v^T\mathbf{h}_i=1$ when the perfect alignment is reached, while such a relation is impossible in Euclidean space: a vector is both parallel to two vectors orthogonal to each other. Therefore, balancing spectrum could be poisonous to alignment of positive pairs and vice versa. In figure \ref{poison}, we investigate how erank, training loss, and accuracy (nDCG@10) change with the intensity of spectrum balancing by adjusting $\alpha$ of DirectSpec. (\textit{i.e.,} larger (smaller) $\alpha$ indicates intenser (weaker) spectrum balancing). We can see that erank rises as increasing $\alpha$, reaching almost $d$ at $\alpha=0.1$. In the meanwhile, the training loss representing the distance among positive pairs (\textit{i.e.,} lower (higher) loss indicates a better (worse) alignment of positive pairs) accordingly rises with erank in Figure \ref{poison} (a), demonstrating the contradiction between spectrum balancing and alignment. The accuracy increases and then drops with $\alpha$, reaching the maximum at $\alpha=0.02$ and erank$=62.9$ in Figure \ref{poison} (b), showing that the perfect balanced spectrum does not necessarily lead to better performance. In conclusion, a good representation lies in the trade-off between alignment and spectrum balancing, rather than focusing solely on either of them.

\subsection{Individualized Spectrum Balancing (DirectSpec$^+$)}
As pointed out in Section 4.2 that balancing the embedding spectrum via DirectSpec is equivalent to the orthogonalization of unobserved pairs with each user/item being optimized as:
\begin{equation}
\mathbf{h}_i\leftarrow \mathbf{h}_i - \sum_j \alpha \left( \mathbf{h}_i^T \mathbf{h}_j \right) \mathbf{h}_j.
\end{equation}
Compared to $\alpha$, $\mathbf{h}_i^T \mathbf{h}_j$ can be considered as an individualized coefficient controlling the intensity for balancing the spectrum between $i$ and $j$. Given that the goal is not solely to generate a perfectly spectrum but rather to balance the trade-off between alignment and spectrum optimization, we argue that DirectSpec does not fully consider the following two factors:
\begin{itemize}[leftmargin=10pt]
\item The intensity should be adaptive to the extent of collapse (\textit{i.e.,} the pairs suffering more (less) from the collapse issue should be balanced with a higher (lower) intensity ).

\item Unobserved pairs may also show relations with varying degrees (\textit{i.e.,} the pairs showing closer relations should be balanced with a lower intensity).

\end{itemize} 
We propose an enhanced individualized spectrum balancing design DirectSpec$^+$ as follows:
\begin{equation}
\begin{aligned}
&\mathbf{H}_{B}^U\leftarrow\mathbf{H}_{B}^U-\alpha \cdot {\rm softmax}\left(\mathbf{H}_{B}^U{\mathbf{H}_{B}^U}^T \odot \mathbf{\Gamma}\right)\mathbf{H}_{B}^U,\\
&\mathbf{H}_{B}^I\leftarrow\mathbf{H}_{B}^I-\alpha \cdot  {\rm softmax}\left(\mathbf{H}_{B}^I{\mathbf{H}_{B}^I}^T \odot \mathbf{\Gamma} \right) \mathbf{H}_{B}^I,\\
\end{aligned}
\label{directspec+}
\end{equation}
where $\odot$ represents the element-wise multiplication. Here, the individualized coefficient between $i$ and $j$ is replaced by $\frac{\exp(\tau_{ij}\cdot\mathbf{h}_i^T \mathbf{h}_j )}{\sum_k \exp(\tau_{ik}\cdot\mathbf{h}_i^T \mathbf{h}_k )}$, where the pairs suffering more from the collapse (\textit{i.e.,} $\mathbf{h}_i^T \mathbf{h}_j\rightarrow1$) are balanced with a higher intensity. $\tau_{ij}$ is a temperature reflecting the relations between $i$ and $j$ estimated from the interaction data, and we propose two adaptive temperature designs to adapt to the pairs with varying degrees of correlations:
\begin{itemize}[leftmargin=10pt]

\item[i.] We dynamically learn the temperature through an attention mechanism such as $\mathbf{\Gamma}_{uv}=\sigma(\mathbf{W}^T[\mathbf{h}_u || \mathbf{h}_v])$ for $u$ and $v$, where $||$ stands for the concatenate operation and $\mathbf{W}\in\mathbb{R}^{2d}$ is the transform matrix.

\item [ii.] We use a unparameterrized polynomial graph filter $\phi (\cdot)=\sum_{l=0}^L \mathbf{\hat{A}}^l$ to measure the relation: $\tau_{uv}=\pi(\cos(\phi(u),\phi(i))$), where $\cos(\cdot,\cdot)$ is the cosine similarity. The user/item pairs with a higher similarity indicating a closer relation should be optimized with a smaller temperature (\textit{i.e.,} lower intensity). Thus, we use a linear function $\pi(\cdot)$ mapping the similarity to a proper range $[\tau_0, \tau_1]$: the highest and lowest similarity is mapped to $\tau_0$ and $\tau_1$, respectively. 
\end{itemize}
We will compare the two designs in Section 5.3.2. We substitute Equation (\ref{directspec+}) into 
Algorithm \ref{directspec} and obtain the running algorithm for DirectSpec$^+$.

\subsection{Discussion}

\subsubsection{Connection with SSL-based Methods}
Self-supervised learning (SSL) has shown great potential in recent years. Some SSL methods can generate high-quality representations without explicitly sampling negative samples that have been applied to recommendation as well, such as Barlow Twins (BT) \cite{zbontar2021barlow}, LogDet \cite{chen2023towards}, and spectral contrastive loss (SCL) \cite{haochen2021provable}:
\begin{equation}
\begin{aligned}
&\mathcal{L}_{BT}=\sum_i \left(1- \mathcal{C}_{ii} \right)^2+ \lambda \sum_i \sum_{j\neq i} \mathcal{C}_{ij}^2, \\
& \mathcal{L}_{logdet}= \Tr \left( \Sigma \right)-\log \det \left( \Sigma \right),\\
& \mathcal{L}_{SCL}= \sum_i \sum_{i\neq j} \left( \mathbf{h}_i^T \mathbf{h}_j \right)^2,
\end{aligned}
\end{equation} 
where $\mathcal{C}_{ij}=\mathbf{H}[:,i]^T\mathbf{H}[:,j]$ is the covariance, and $\Sigma=\mathbf{H}^T\mathbf{H}$ is the covariance matrix.
\begin{proposition}
BT, LogDet, and SCL are special cases of DirectSpec (Equation (\ref{balance_spec})) with $f\left( \sigma_1,\cdots,\sigma_d  \right)=\sum_i (\sigma_i^2-1)^2$ for BT and SCL and $f\left( \sigma_1,\cdots,\sigma_d  \right)=\sum_i \sigma_i- 2\log \sigma_i$ for LogDet, respectively.
\end{proposition}
With $\mathbf{H}$ being normalized column-wisely, $\sum \sigma_i^2=\Tr(\mathbf{Q}diag(\sigma_i^2)\mathbf{Q}^T)=\Tr(\mathbf{H}^T\mathbf{H})=d$. Then, Barlow Twins can be rewritten as follows:
\begin{equation}
\begin{aligned}
\mathcal{L}_{BT}&=\Tr\left(\left(\mathbf{H}^T\mathbf{H}-\mathbf{I} \right)^2 \right)= \Tr\left( \mathbf{Q}diag\left( \sigma_i^2 \right)\mathbf{Q}^T-2\mathbf{Q}diag\left( \sigma_i \right)\mathbf{Q}^T+ \mathbf{Q}\mathbf{Q}^T\right)\\
&=\Tr\left( \mathbf{Q}diag\left( (\sigma_i^2-1)^2 \right)\mathbf{Q}^T\right)
=\sum_i \left(\sigma_i^2-1\right)^2.
\label{directspec_barlow}
\end{aligned}
\end{equation} 
With $\mathbf{H}$ being normalized row-wisely, we can rewrite SCL as a similar form to BT:
\begin{equation}
\mathcal{L}_{SCL}=\Tr\left(\left(\mathbf{H}\mathbf{H}^T-\mathbf{I} \right)^2 \right)=\Tr\left( \mathbf{P}diag\left( (\sigma_i^2-1)^2 \right)\mathbf{P}^T\right)=\sum_i \left(\sigma_i^2-1\right)^2.
\label{directspec_scl}
\end{equation} 
Similarly, we can reformulate LogDet as:
\begin{equation}
\mathcal{L}_{logdet}=\Tr\left(\mathbf{H}^T\mathbf{H}\right)-\log \det \left(\mathbf{H}^T\mathbf{H} \right)=\sum_i \sigma_i^2 - \log \prod \sigma_i^2=\sum_i \sigma_i^2-2\log \sigma_i.
\label{directspec_logdet}
\end{equation}
By substitute Equation (\ref{directspec_barlow}) (or (\ref{directspec_scl})) and (\ref{directspec_logdet}) into Equation (\ref{balance_spec}) and solve them, we can get the minima achieved at $\sigma_1=\cdots=\sigma_d=\sqrt{d}$ for the three methods, $\beta=1-2d$ and  $\beta=\frac{1-d}{d}$ for BT/SCL and LogDet when $c=d$, respectively, demonstrating that they can prevent the representations from collapsing by implicitly balancing the embedding spectrum. Since they belong to the DirectSpec framework, our proposed DirectSpec$^+$ balances spectrum more adaptively and effectively, showing superiority over them. It is worth noting that InfoNCE loss cannot be reformulated as a case of DirectSpec. However, the embeddings are nearly and tend to be more orthogonal to each other as the dimensionality $d$ becomes larger, indicating a perfectly balanced spectrum, as shown in Proposition 3 of \cite{wang2020understanding}. We will compare our proposed methods with them in Section 5. Given DirectSpec's generality and similarity to SSL methods that have been applied across various domains, we believe that DirectSpec can also contribute to other domains such as images, graph-structured data, etc.

\begin{table}
\caption{Comparison of time complexity.}
\scalebox{1.1}{
\begin{tabular}{c|c|c|c|c}
\toprule
\textbf{Model} & DirectSpec  &Barlow Twins &SCL  &LogDet                         \\ \hline
\textbf{Complexity} & $c|\mathbf{R}^+|Bd$ & $c|\mathbf{R}^+|d^2$ &$c|\mathbf{R}^+|Bd$& $c|\mathbf{R}^+|d^2\left( 1+ \frac{d}{B} \right)$ \\ \bottomrule
\end{tabular}}
\label{complexity}
\end{table}

\begin{table}
\centering
\caption{Statistics of datasets.}
\scalebox{1.15}{
\begin{tabular}{lcccc}
\toprule
Datasets&\#User&\#Item &\#Interactions &Density\%\\
\midrule
CiteULike&5,551&16,981&210,537&0.223\\
Yelp&25,677&25,815&731,672&0.109\\
Gowalla&29,858&40,981&1,027,370&0.084\\
\bottomrule
\label{datasets}
\end{tabular}}
\end{table}
   
\subsubsection{Complexity}
We compare the complexity of DirectSpec$^+$ with Barlow Twins, spectral contrastive loss, and LetDet in Table \ref{complexity}, where $B$ and $c$ denote the batch size and the required epochs, respectively. We ignore the complexity for optimizing the alignment here. Overall, LogDet has the most expensive computational cost as it requires computing the matrix determinant with complexity as $\mathcal{O}(d^3)$. The complexity of DirectSpec$^+$ is identical to that of SCL.
The difference of complexity between DirectSpec$^+$ and BT lies in the term $Bd$ and $d^2$, where DirectSpec$^+$ might be more costly when $B$ is large. However, considering that we set $B=256$ in the experiments, and the complexity of back propagation for BT and SCL is still $c|\mathbf{R}^+|Bd$ which is not contained in Table \ref{complexity}, DirectSpec$^+$ is more time-efficient than BT and SCL. 

\subsubsection{Comparison between Embedding Collapse and Over-Smoothing}
Generally, embedding collapse can be mathematically formulated as follows:
\begin{equation}
{\rm rank} \left( \mathbf{H}^{(L)} \right) < \epsilon,
\label{collapse_math}
\end{equation} 
where $ \mathbf{H}^{(L)}$ is the embedding when erank converges after updating $L$ times, $\epsilon<d$, and a smaller $\epsilon$ indicates more serious embedding collapse. Over-smoothing is attributed to the shrunk graph spectrum as repeating the message passing that can also be formulated as Equation~(\ref{collapse_math}). The collapse issue analyzed in this work shares similarities with over-smoothing, due to the observation that the parameter-updating process in recommendation algorithm shares the same form as the message passing in GNN. However, since the cause of over-smoothing in GNN lies in the graph spectrum, existing works tackle this issue mostly by dealing with the graph spectrum, such as randomly dropping out the edges to perturb the spectrum \cite{rong2020dropedge}, balancing the contribution of different order neighborhood \cite{gasteiger2018predict,peng2022less} to avoid all eigenvalues from shrinking, etc., which fail to solve the embedding collapse in this work, as it exists generally on recommendation algorithms without relying on explicit graph structures. On the other hand, our proposed DirectSpec tackles a general embedding collapse by directly balancing the embedding spectrum which is proved to solve this issue. We will show that our DirectSpec is also effective on over-smoothing and embedding collapse caused by other reasons in Section 5.  

\subsubsection{Comparison with Related Works}
Recently, several efforts have been made to address `collapse issues' in recommendation methods  \cite{chen2023towards,zhang2023mitigating,guo2023embedding}, which shares similarity with our work. Guo et al.~\cite{guo2023contranorm} points out an embedding collapse within the feature interaction module for click-rate prediction (CTR) caused by the imbalanced gradients over different singular vectors, while the embedding collapse issue in our work stems from optimizing a CF algorithm. The dimensional collapse in graph collaborative filtering (GCF) investigated in \cite{chen2023towards,zhang2023mitigating} overlap more significantly with ours. Both the works exploit a LogDet function to address the issue, which are more computationally expensive than DirectSpec as analyzed in Section 4.4.2. We will demonstrate the superiority of our proposed methods in Section 5.

\section{Experiments}
In this section, we comprehensively evaluate our proposed methods on three public datasets. We implement DirectSpec and DirectSpec$^+$ on two popular baselines: MF and LightGCN. Particularly, we introduce data description and implementation details in Section 5.1; we compare the proposed methods with other competitive baselines in terms of accuracy and efficiency in Section 5.2. Finally, we conduct model analysis in Section 5.3, including detailed analysis and experimental results on how DirectSpec$^+$ prevents embeddings from collapsing without relying on negative samples, and how different settings of hyperparameters affect model performance. 

\subsection{Experimental Settings}
\subsubsection{Datasets} We use the following three publicly available real-world datasets in this work, where the statistics of them are summarized in Table \ref{datasets}.
\begin{itemize}[leftmargin=10pt]
\item \textbf{CiteULike}\footnote{\url{https://github.com/js05212/citeulike-a}}: This is an scholarly article recommendation dataset. Users are allowed to create their own collections of articles including abstracts, titles, and tags, etc.  
\item \textbf{Yelp} \cite{he2016fast}: This is a business dataset from Yelp Challenge data. The items are point of interests (POIs), users can leave reviews and ratings.
\item \textbf{Gowalla} \cite{wang2019neural}: This is a check-in datasets recording which locations users have visited. 
\end{itemize}   
Since we focus on CF for implicit feedbacks, we remove auxiliary information such as reviews, tags, geological and item information, etc, and transform explicit ratings to 0-1 implicit feedbacks.

\subsubsection{Evaluation Metrics}
We adopt two widely used evaluation metrics for personalized rankings: Recall and normalized discounted cumulative gain (nDCG) \cite{jarvelin2002cumulated} to evaluate model performance. The recommendation list is generated by ranking unobserved items and truncating at position $k$. Recall measures the ratio of the relevant items in the recommendation list to all relevant items in test sets, while nDCG takes the rank into consideration by assigning higher scores to the relevant items ranked higher. We use 80\% of the interactions for training and randomly select 10\% from the training set as validation set for hyper-parameter tuning, the rest 20\% data is used for test sets; we report the average accuracy on test sets.

\subsubsection{Baselines}
We compare DirectSpec with the following competitive baselines:
\begin{itemize}[leftmargin=10pt]

\item BPR \cite{rendle2009bpr}: This is a stable and classic MF-based method, exploiting a Bayesian personalized ranking loss for personalized rankings.

\item ENMF \cite{chen2020efficient}: This is a neural recommendation model performing whole-data-based learning without sampling. We choose user-based ENMF as the baseline. 

\item CCL \cite{mao2021simplex}: The proposed cosine contrastive loss (CCL) maximizes the similarity between positive pairs and minimizes the similarity of negative pairs below the margin $m$. After parameter tuning, we set $\mathcal{N}=1000$, $w=300$ on all datasets, $m=0.1$, $0.3$, and $0.3$ on CiteULike, Yelp, and Gowalla, respectively, and we choose MF as the encoder. 

\item DirectAU \cite{wang2022towards}: This method directly optimizes alignment and uniformity and shows superior performance. Following the original paper, we choose MF and LightGCN as the encoder, and set $\gamma=5.0$, 0.5, and 1.5 on CiteULike, Yelp, and Gowalla, respectively.

\item LightGCN \cite{he2020lightgcn}: This is a linear GNN method that only keeps neighborhood aggregation for recommendation. We employ a three-layer architecture as our baseline.

\item SGL-ED \cite{wu2020self}: This model explores self-supervised learning by maximizing the agreements of multiple views from the same node, where the node views are generated by adding noise such as randomly removing the edges or nodes on the original graph. We set $\tau=0.2$, $\lambda_1=0.1$, $p=0.1$, and use a three-layer architecture.

\item LightGCL \cite{caisimple}: This is a simple yet effective graph contrastive
learning method injecting global collaborative relations via singular value decomposition. 

\item GDE \cite{peng2022less}: This method only keeps a very few graph features for recommendation without stacking layers, showing less complexity and higher efficiency than GNN-based methods. 

\item LogDet \cite{zhang2023mitigating}: This is a decorrelation-enhanced method to mitigate the dimensional collapse of graph collaborative filtering.

\item AFDGCF \cite{wu2024afdgcf}: This method tackles the over-correlation issue in GNN-based methods, and shares significant similarities with Barlow Twin \cite{zbontar2021barlow}. It can be considered as the implementation of Barlow Twin for recommendation. We choose LightGCN as the encoder.

\end{itemize} 

\subsubsection{Implementation Details}
We implemented our DirectSpec based on PyTorch, and the code is released publicly\footnote{https://github.com/tanatosuu/directspec}. SGD is adopted as the optimizer for all models, the embedding size $d$ is set to 64, the regularization rate is set to 0.01 on all datasets, the learning rate is tuned amongst $\{0.001,0.005,0.01,\cdots\}$; without specification, the model parameters are initialized with Xavier Initialization \cite{glorot2010understanding}; the batch size is set to 256. We report other hyperparameter settings in the next subsection.

\begin{table}[]
\caption{Performance comparison. The best baseline is underlined. ``$*$'' indicates statistical significance at $p<0.01$ for a one-tailed t-test. }
\scalebox{0.9}{
\begin{tabular}{cccccc}
\toprule
\textbf{Datasets}                    & \textbf{Methods} & \textbf{nDCG@10}               & \textbf{nDCG@20}               & \textbf{Recall@10}             & \textbf{Recall@20}             \\ \hline
\multirow{11}{*}{\textbf{CiteULike}} & BPR              & 0.1620                         & 0.1773                         & 0.1778                         & 0.2190                         \\                              
                                     & CCL(MF)           & 0.1545                         & 0.1642                         & 0.1716                         & 0.1996 \\
                                     
                                     & ENMF-U           &0.1335							 &0.1601 &0.1556 						&0.2193 \\                                     
                                     
                                     & DirectAU(MF)         & \underline{0.2102}                   & \underline{0.2252}                   & \underline{0.2260}                   & \underline{0.2693}                   \\
                                      & DirectAU(GCN)         & 0.1926 & 0.2109 & 0.2116                  & 0.2604                   \\                                    
                                     & LightGCN         & 0.1610                         & 0.1781                         & 0.1771                         & 0.2190                         \\
                                     & SGL-ED           & 0.1890                         & 0.2065                         & 0.2117                         & 0.2588                         \\
                                     & LightGCL         &0.2096                          &0.2238 &0.2214                          &0.2638                         \\                                     
                                     & GDE              & 0.1890                         & 0.2061                         & 0.2055                         & 0.2528                         \\
                                     &LogDet           & 0.1403            & 0.1579            & 0.1548              & 0.2034         						 \\
                                     &AFDGCF           & 0.1619            & 0.1787            & 0.1790              & 0.2246         						 \\                                     
                                     
                                     & DirectSpec(MF)   & 0.1688                         & 0.1849                         & 0.1827                         & 0.2270                         \\
                                     & DirectSpec(GCN)  & 0.1693                         & 0.1863                         & 0.1875                         & 0.2334                         \\
                                     & DirectSpec$^+$(MF)  & \textbf{0.2197}$^*$                & \textbf{0.2354}$^*$                & \textbf{0.2352}$^*$                & \textbf{0.2818}$^*$ \\
                                     & DirectSpec$^+$(GCN) & 0.2038                         & 0.2213                         & 0.2213                         & 0.2704                         \\                            \hline
\multirow{11}{*}{\textbf{Yelp}}      & BPR              & 0.0487                         & 0.0583                         & 0.0607                         & 0.0869                         \\                                                                       
                                     & CCF(MF)           & 0.0509                         & 0.0593 & 0.0617                         & 0.0846                         \\

                                     & ENMF-U           &0.0665							 &0.0811 &0.0786 						&0.1167 \\                                       
                                     
                                     & DirectAU(MF)         & \underline{0.0721 }                  & \underline{0.0848}                   & \underline{0.0872}                   & \underline{0.1219}                   \\
                                     & DirectAU(GCN)         & 0.0695 & 0.0819 & 0.0854                  & 0.1194                   \\
                                     & LightGCN         & 0.0572                         & 0.0690                         & 0.0721                         & 0.1045                         \\
                                     & SGL-ED           & 0.0676                         & 0.0794                         & 0.0837                         & 0.1166                         \\
                                     & LightGCL          &0.0673            &0.0790      &0.0836                                     &0.1151                       \\                                     
                                     & GDE              & 0.0653                         & 0.0771                         & 0.0805                         & 0.1129                         \\
                                     &LogDet           & 0.0711            & 0.0833            & 0.0863              & 0.1203                                                        \\                                     
                                     &AFDGCF           & 0.0585            & 0.0702            & 0.0743              & 0.1065         						 \\                                     
                                     & DirectSpec(MF)   & 0.0689                         & 0.0804                         & 0.0839                         & 0.1156                         \\
                                     & DirectSpec(GCN)  & 0.0712                         & 0.0832                         & 0.0861                         & 0.1192                         \\
                                     & DirectSpec$^+$(MF)  & 0.0743                         & 0.0864                         & \textbf{0.0909}$^*$                          & 0.1249 \\
                                     & DirectSpec$^+$(GCN) & \textbf{0.0745}$^*$                & \textbf{0.0868}$^*$                & \textbf{0.0909}$^*$                & \textbf{0.1252}$^*$                \\                                     \hline
\multirow{11}{*}{\textbf{Gowalla}}   & BPR              & 0.1164                         & 0.1255                         & 0.1186                         & 0.1483                         \\
                                 
                                                                          & CCL(MF)           & 0.1269                         & 0.1349                         & 0.1295                         & 0.1573                         \\
                                                                          
                                                                          & ENMF-U           & 0.1166                         & 0.1286                         & 0.1173                         & 0.1528                         \\                                                                          
                                                                          
                                     & DirectAU(MF)         & 0.1286                         & 0.1402                         & 0.1349                         & 0.1710                         \\
                                     & DirectAU(GCN)         &0.1298 &0.1409 &0.1363& 0.1711                   \\
                                     & LightGCN         & 0.0987                         & 0.1098                         & 0.1074                         & 0.1399                         \\
                                     & SGL-ED           & 0.1343                   & 0.1462                   & 0.1417                   & 0.1779                  \\
                                     & LightGCL           & \underline{0.1368} & \underline{0.1482} & \underline{0.1436}                         & \underline{0.1801} \\                                                         
                                     & GDE              & 0.1261                         & 0.1367                         & 0.1313                         & 0.1656                         \\
                                     & LogDet           & 0.1179            & 0.1272            & 0.1203             & 0.1511                           \\  
                                      &AFDGCF           & 0.1011            & 0.1117            & 0.1087              & 0.1406         						 \\                                                                   
                                     & DirectSpec(MF)   & 0.1133                         & 0.1225                         & 0.1183                         & 0.1480                         \\
                                     & DirectSpec(GCN)  & 0.1160                         & 0.1251                         & 0.1191                         & 0.1490                         \\
                                     & DirectSpec$^+$(MF)  & 0.1389                & 0.1509                         & 0.1447               & 0.1819                         \\
                                     & DirectSpec$^+$(GCN) & \textbf{0.1405}$^*$                         & \textbf{0.1518}$^*$                 & \textbf{0.1467}$^*$                 & \textbf{0.1837}$^*$                                        \\                                     \bottomrule
\end{tabular}}
\label{performance_comparison}
\end{table}

\begin{table}[]
\caption{Training time (seconds) per epoch.}
\begin{tabular}{c|c|c|c}
\toprule
Model/Data      & Citeulike    & Yelp          & Gowalla       \\ \hline
BPR             & 2.37         & 8.53          & 13.28         \\
DirectSpec$^+$(MF)  & 3.78 (1.59x) & 13.15 (1.54x) & 19.10 (1.44x) \\
LigtGCN         & 8.78         & 43.18         & 79.79         \\
DirectSpec$^+$(GCN) & 9.49 (1.08x) & 45.70 (1.06x) & 81.20 (1.02x) \\ 
SGL-ED(GCN) & 22.20 (2.53x) & 182.54 (4.23x) & 401.96 (5.04x) \\ 
LightGCL(GCN) & 11.59 (1.32x) & 51.38 (1.19x) & 92.56 (1.16x) \\ 
AFDGCF(GCN) & 11.45 (1.30x) & 48.53 (1.12x) & 84.86 (1.06x) \\ 
\bottomrule
\end{tabular}
\label{efficieny_comparison}
\end{table}

\subsection{Comparison}
\subsubsection{Performance}
\begin{figure} \centering 
\subfigure {  
\includegraphics[width=0.43\columnwidth]{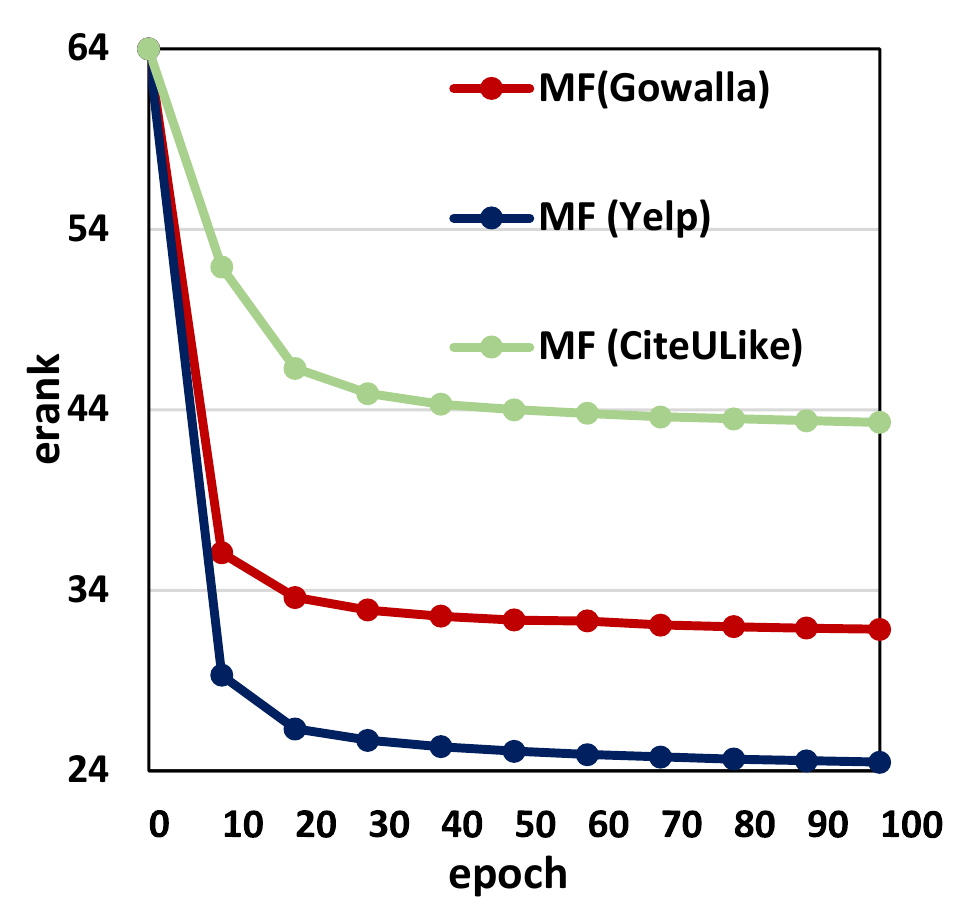} 
}
\subfigure {  
\includegraphics[width=0.43\columnwidth]{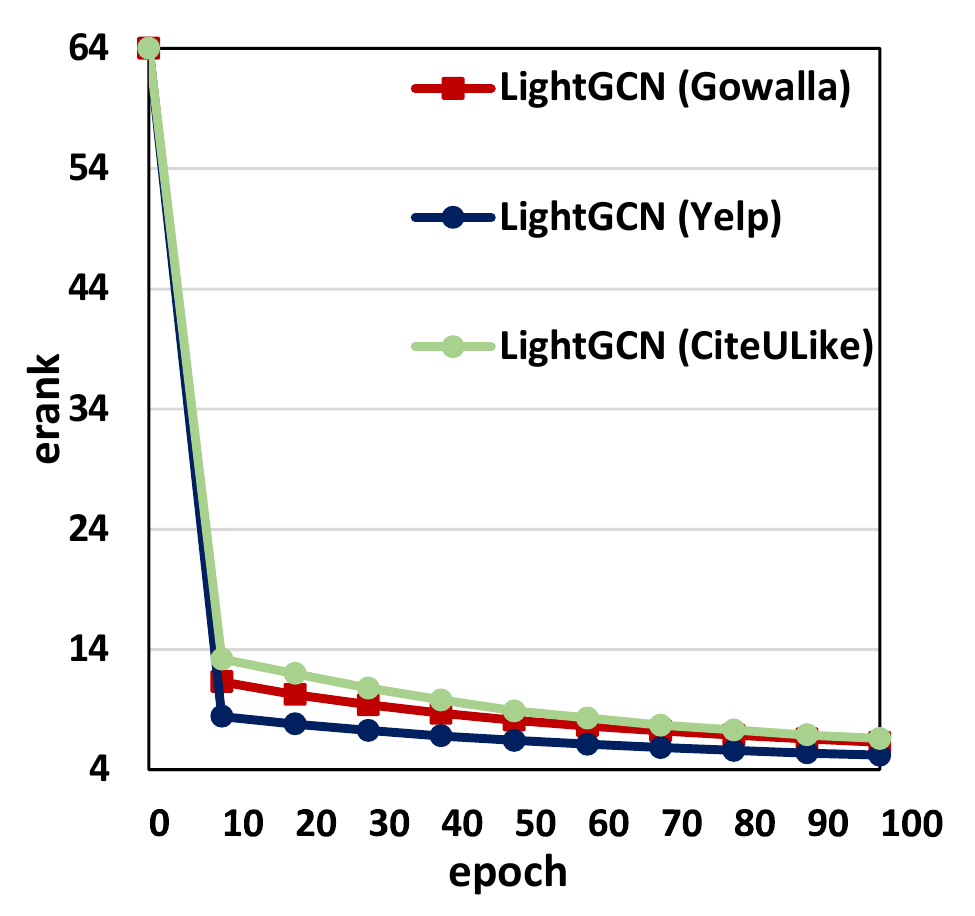} 
}
\caption{The extent of collapse on three datasets.}           
\label{data_collapse}
\end{figure}

We report overall performance in Table \ref{performance_comparison}, and observe the followings:
 
\begin{itemize}[leftmargin=10pt]
\item Overall, GNN-based methods show better performance on sparse data (Yelp and Gowalla) than dense data (CiteULike). For instance, DirectSpec$^+$(GCN) achieves better and worse than DirectSpec$^+$(MF) on Gowalla and CiteULike, respectively, and their performance is close on Yelp. LightGCN underperforms BPR on CiteULike and Gowalla, which might be attributed to the slow convergence that has been reported in the original paper. Among baselines, DirectAU achieves the best on CiteULike and Yelp, while LightGCL outperforms other baselines on Gowalla. Our proposed DirectSpec$^+$ implemented on both LightGCN and MF consistently show improvements over all baselines.

\item ENMF and CCL exploit enhanced negative sampling for training, while the superior performance over them demonstrates the superiority of DirectSpec learning paradigm.

\item Unlike negative sampling, SSL-based methods adopt different training strategies to enhance the data representations. Amongst them, DirectAU directly regulating the uniformity shows relatively superior performance, while AGDGCF utilizing a Barlow-Twin-like model to remove the feature correlation shows poor performance.

\item Since our methods are implemented on MF and LightGCN, the effectiveness of our methods can be further demonstrated by directly comparing with them. DirectSpec(MF) and DirectSpec$^+$(MF) outperform BPR by 14.3\% and 35.8\%, on average in terms of nDCG@10, respectively. In the meanwhile, the improvement of DirectSpec(GCN) and DirectSpec$^+$(GCN) over LightGCN is 15.7\% and 33.1\% on average in terms of nDCG@10, respectively. 

\item Overall, the improvement of DirectSpec over MF and LightGCN is Yelp$>$Gowalla$>$CiteULike (33.0\%, 7.4\%, and 4.7\% on average in terms of nDCG@10, respectively). Figure \ref{data_collapse} compares the extent of collapse on three datasets using MF and LightGCN (with only positive samples). We can observe that Yelp suffers more from the collapse while CiteULike suffers less, which is consistent with the improvement of DirectSpec. This observation reveals that the dataset suffering more from collapse tends to benefit more from DirectSpec.

\item DirectSpec$^+$ shows significant improvement over DirectSpec across all datasets. As introduced in Section 4.3, we adopt a individualized spectrum balancing policy for DirectSpec$^+$ 
to achieve a better trade-off between alignment and spectrum balancing. The superior performance demonstrates its effectiveness.

\end{itemize}

\begin{figure} \centering 
\subfigure[CiteULike] {  
\includegraphics[width=0.42\columnwidth]{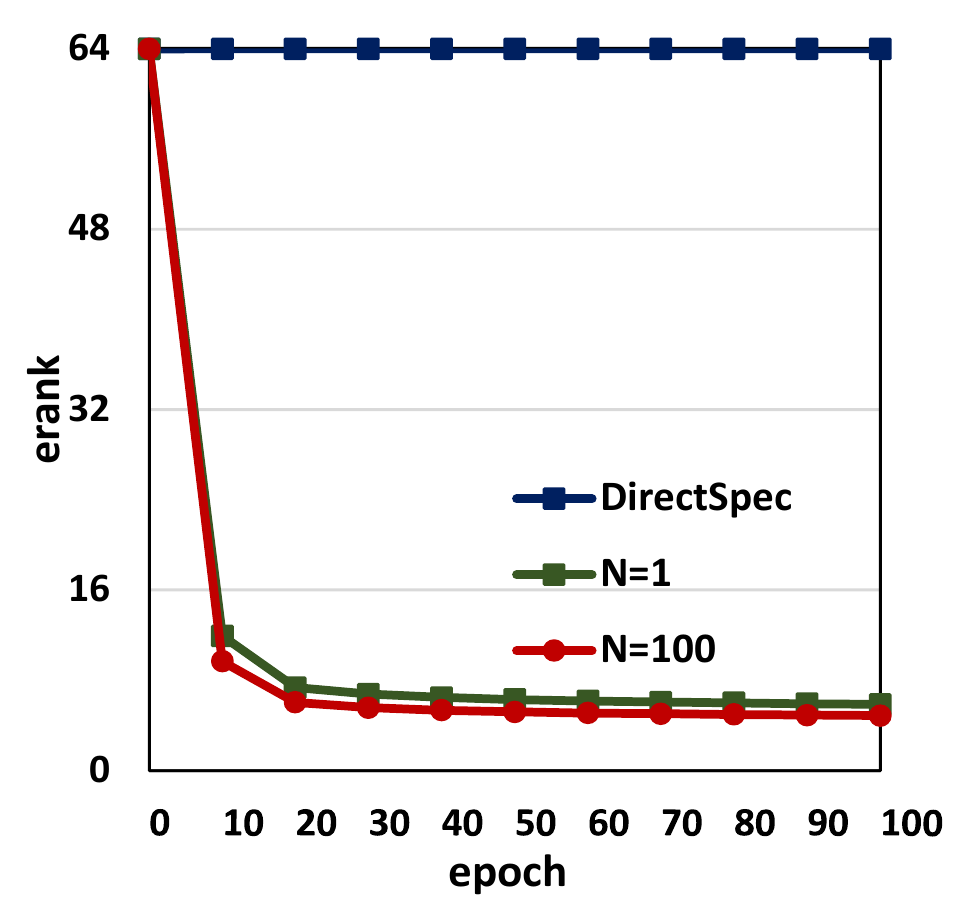} 
}
\hspace{0.2cm}
\subfigure[Yelp] {  
\includegraphics[width=0.42\columnwidth]{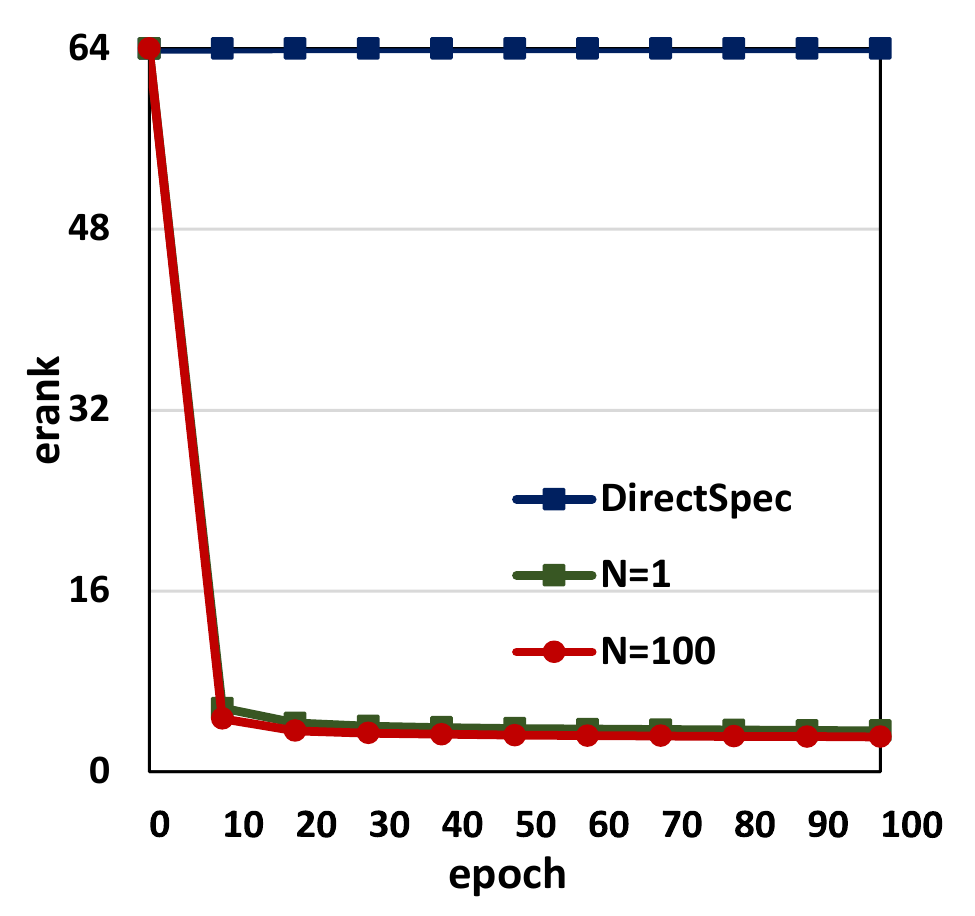} 
}
\hspace{0.2cm}
\subfigure[CiteULike] {  
\includegraphics[width=0.42\columnwidth]{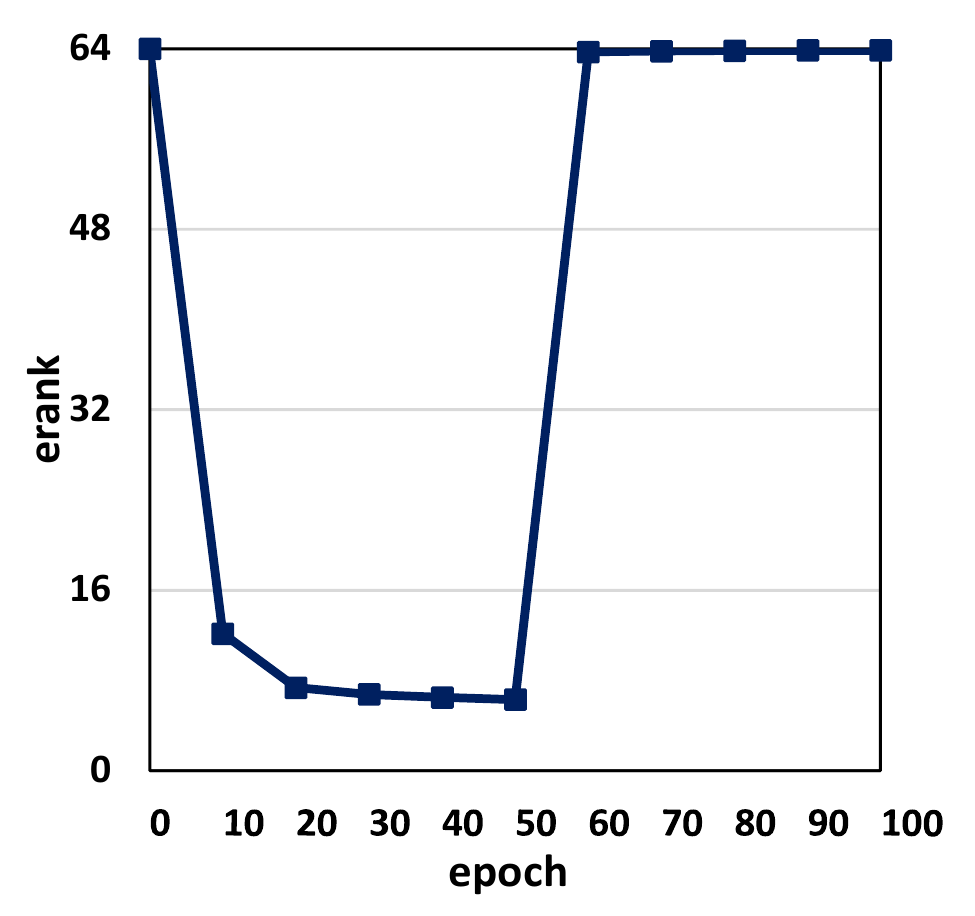} 
}
\hspace{0.2cm}
\subfigure[Yelp] {  
\includegraphics[width=0.42\columnwidth]{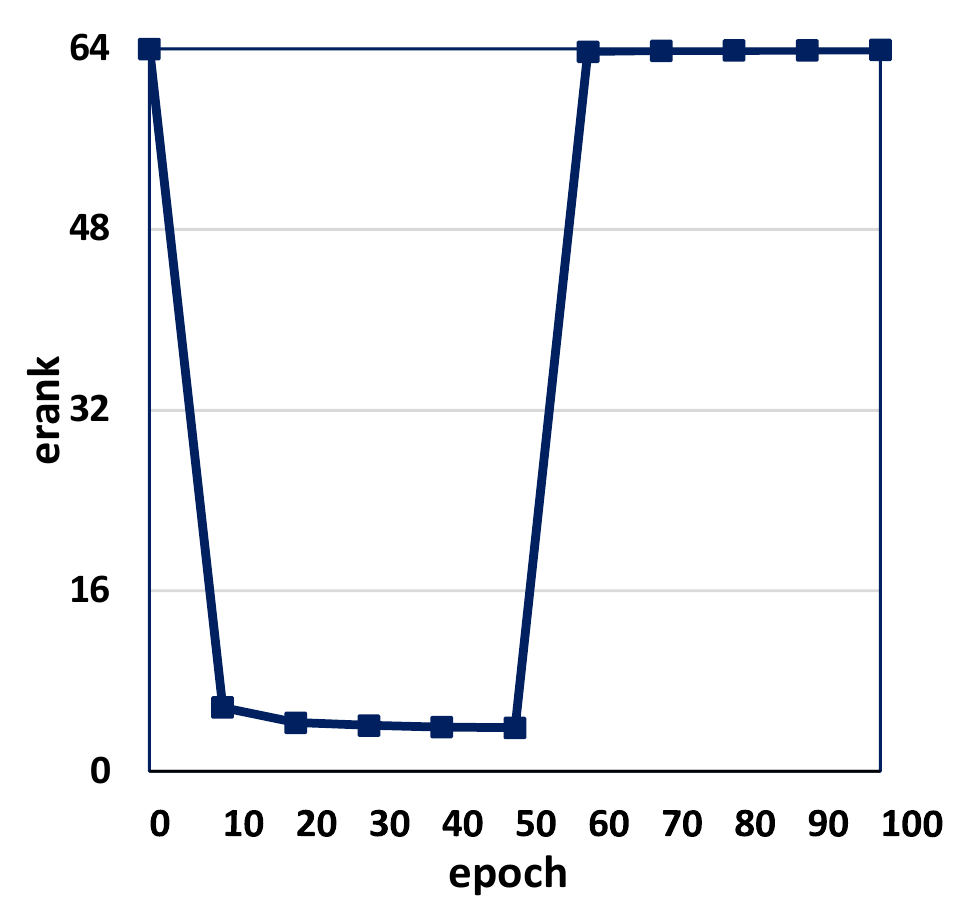} 
}
\caption{All models are optimized without using observed interactions. In (a) and (b), we show how erank changes on DirectSpec (MF) and MF (BCE) where $N$ is the negative sampling ratio. In (c) and (d), we use BCE (MF) for the first 50 epochs and DirectSpec for the last 50 epochs. }           
\label{negative_direct_compare}
\end{figure}

\subsubsection{Efficiency}
We report the training time per epoch of several methods implemented on BPR and LightGCN in Table \ref{efficieny_comparison}. The experiments are all conducted on Intel(R) Core(TM) i9-10980XE CPU and NVIDIA RTX A6000 GPU. We first compare BPR and DirectSpec$^+$(MF), and observe that it takes roughly 0.5x additional running time, which is acceptable considering the significant improvement. Compared to other methods implemented on LightGCN, DirectSpec$^+$ almost shows no additional running time, due to the reason that the complexity of DirectSpec$^+$ is mainly determined by the batch size rather than the size of users and items. Since LightGCN has a higher complexity than BPR, the proportion of DirectSpec$^+$(LightGCN)'s running time per epoch to that of LightGCN is much smaller than the proportion of DirectSpec$^+$(MF)'s running time to that of BPR. This finding can explain another observation that the additional time is smaller on larger datasets (\textit{i.e.,} Gowalla$<$Yelp$<$CiteULike). The complexity of DirectSpec$^+$ is unchanged while LightGCN and BPR accordingly take more running time on larger datasets, thus the proportion of DirectSpec$^+$'s running time to that of LightGCN/BPR becomes smaller. Meanwhile, SGL-ED is more computationally expensive on larger datasets, and DirectSpec$^+$(LightGCN) is more efficient than it with more significant improvement over LightGCN. Furthermore, despite the additional training time per epoch added to the encoder, the scalability and utility have actually been improved. For instance, the required epochs are 250 and 600 for BPR and LightGCN on Gowalla, respectively, as opposed to just 80 and 60 epochs for DirectSpec(MF) and DirectSpec(LightGCN), resulting in 2.2x and 9.8x speed-up, respectively.

\begin{figure} \centering 
\subfigure[AFDGCF on CiteULike] {  
\includegraphics[width=0.31\columnwidth]{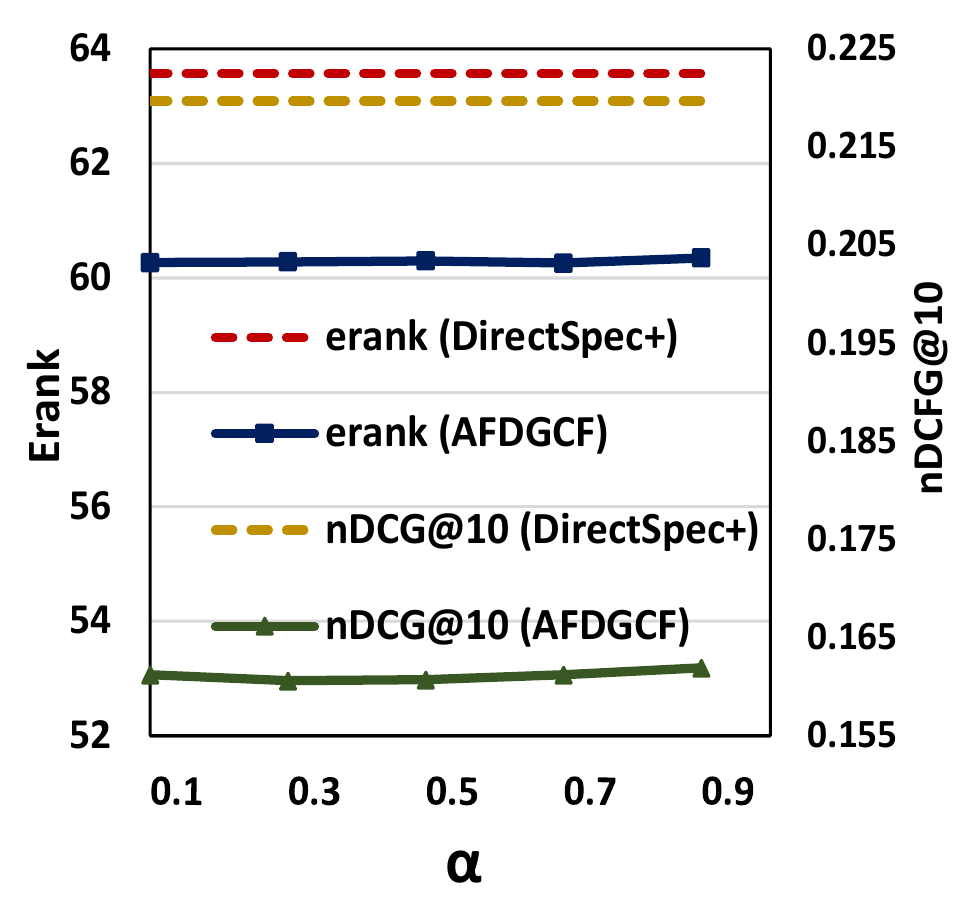} 
}
\subfigure[LogDet on Gowalla] {  
\includegraphics[width=0.31\columnwidth]{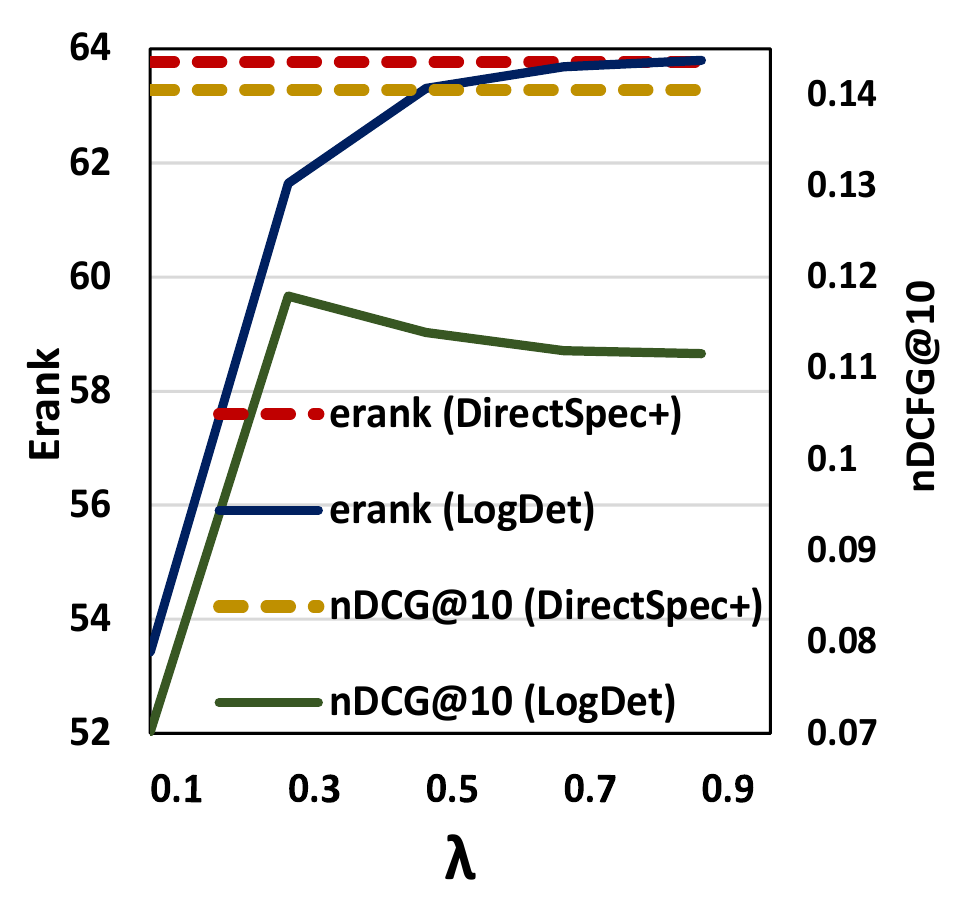} 
}
\subfigure[DropEdge on Yelp] {  
\includegraphics[width=0.31\columnwidth]{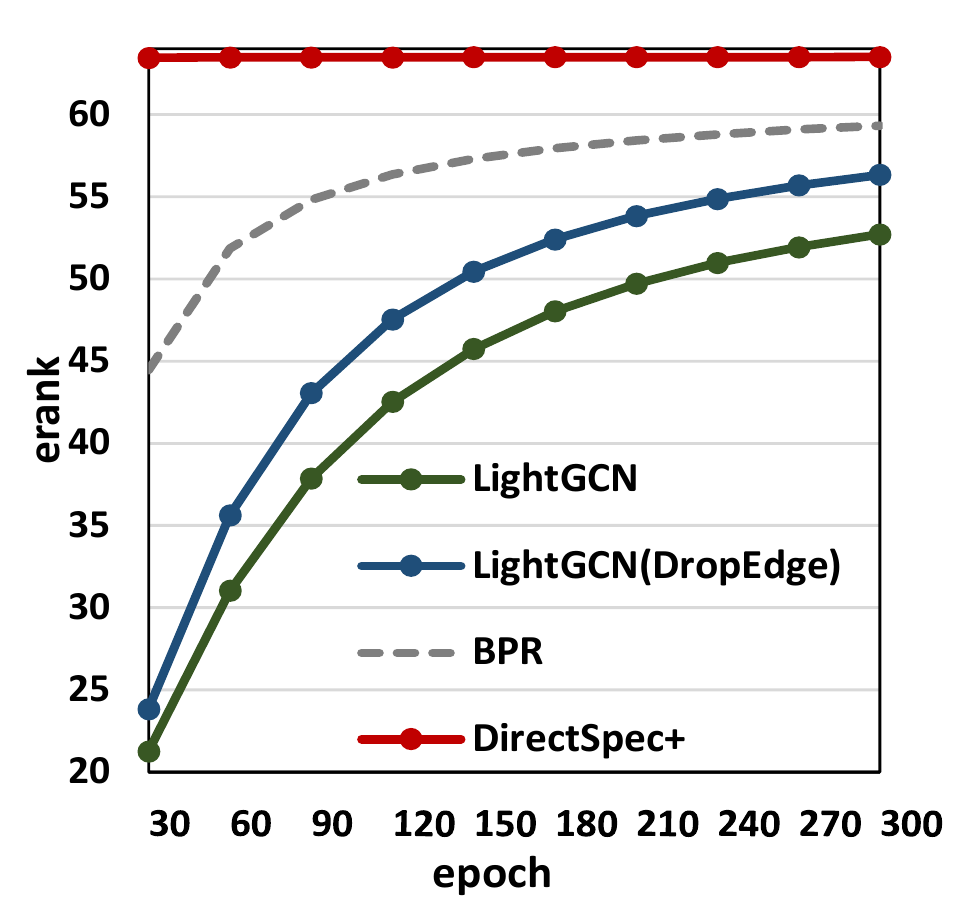} 
}
\caption{Comparison with several methods.}           
\label{other_methods_compare}
\end{figure} 

\begin{figure} \centering 
\subfigure[Erank comparison] {  
\includegraphics[width=0.31\columnwidth]{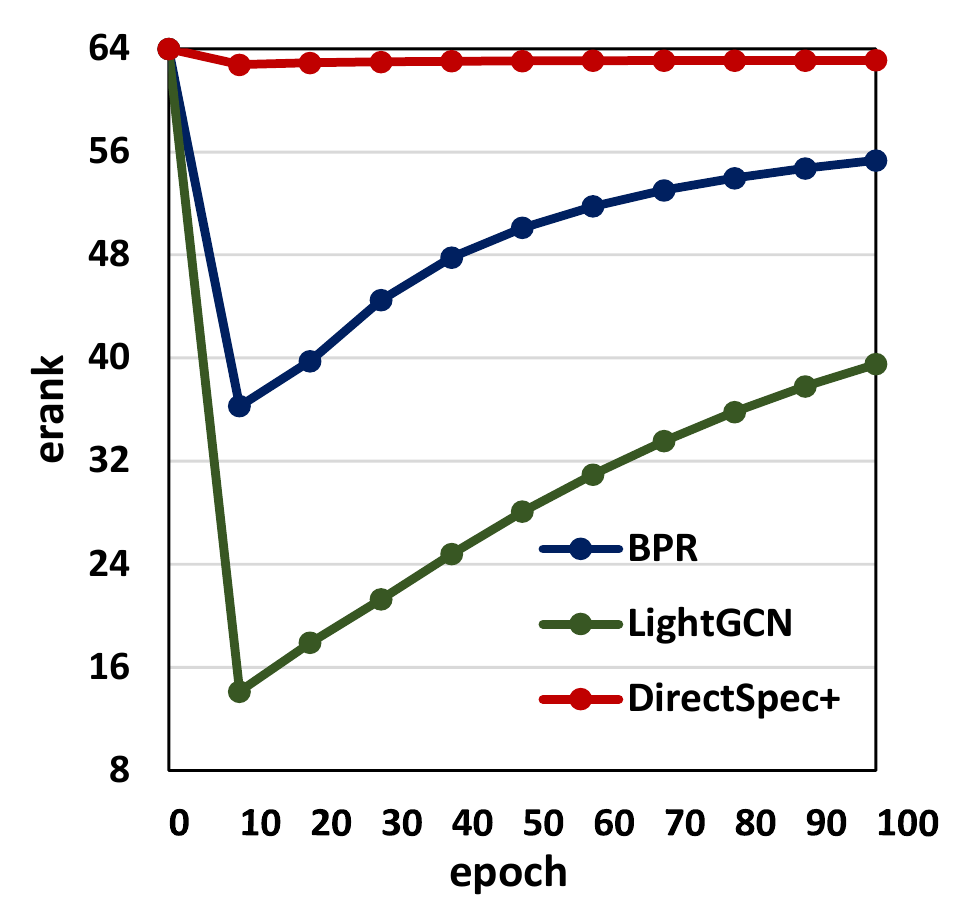} 
}
\subfigure[Singular value distribution of the embedding matrix] {  
\includegraphics[width=0.31\columnwidth]{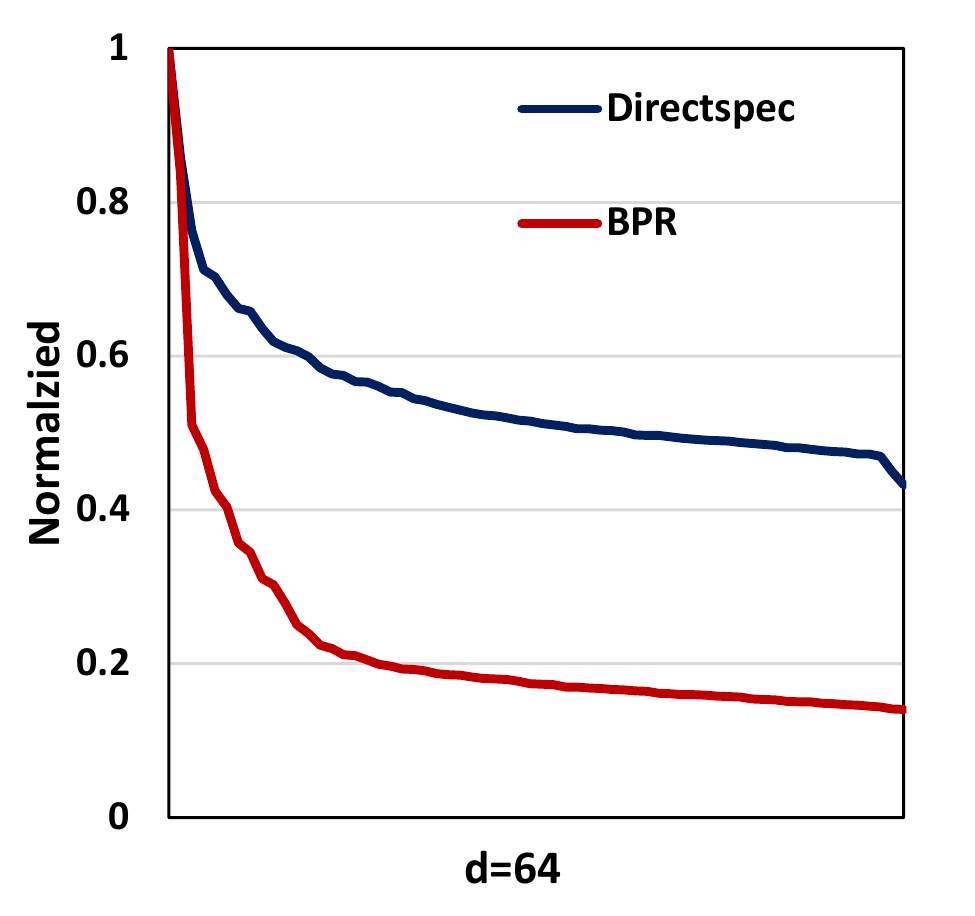} 
}
\subfigure[Comparison of erank and average similarity on pairs in test sets] {  
\includegraphics[width=0.31\columnwidth]{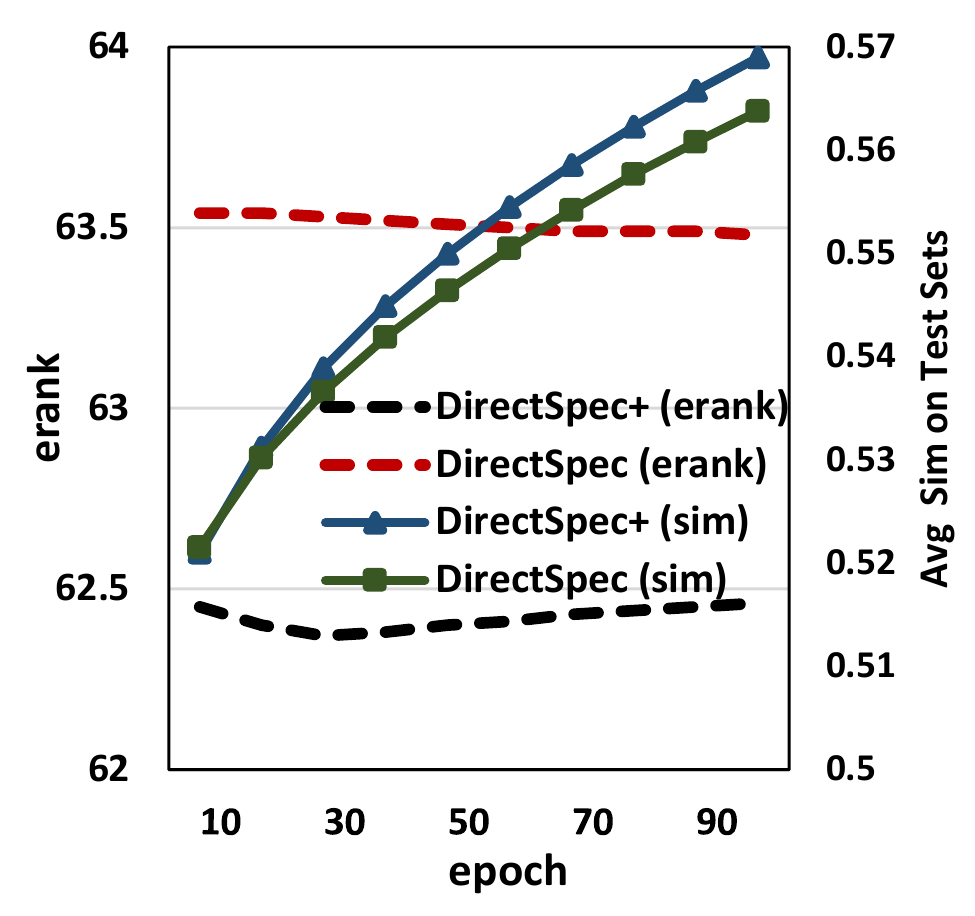} 
}
\caption{How DirectSpec prevents collapse (on Yelp).}           
\label{prevent_collapse}
\end{figure}

\subsubsection{Comparison with Negative Sampling}
The results in Table \ref{performance_comparison} demonstrates the superiority of DirectSpec over negative sampling based methods including BPR and CCL in terms of performance. Furthermore, we compare how these two types of methods alleviate collapse issue by optimizing them solely based on negative pairs. In Figure \ref{negative_direct_compare} (a) and (b), we observe that the erank of MF (BCE) consistently drops without optimization on the positive pairs, resulting in a complete collapse, and the erank drops more quickly as increasing the negative sampling ratios, whereas the erank of DirectSpec remains unchanged throughout the training. This observation indicates that there exists a dynamic balance between positive and negative samplings, the lack of either of them cannot prevent the embeddings from collapsing. Furthermore, since negative sampling is equivalent to a high pass filter, it can only alleviate the collapse caused by low pass filters (\textit{e.g.,} optimization on observed interactions), while DirectSpec is equivalent to an all pass filter that can tackle the collapse under any situations. Figure \ref{negative_direct_compare} (c) and (d) demonstrate that DirectSpec can recover the embeddings from collapsing caused by negative sampling.

\subsubsection{Comparison with SSL-based methods}
Particularly, we focus on AFDGCF (BT) and LogDet that can be considered as special cases of DirectSpec. We show how accuracy and erank change with the hyperparameters $\alpha$ and $\lambda$ controlling the intensity of spectrum balancing in Figure \ref{other_methods_compare} (a) and (b), respectively. We observe that both performance and erank are not sensitive to $\alpha$ on AFDGCF, which are significantly inferior to DirectSpec$^+$. While the erank consistently increases with $\lambda$ and reaches almost $d$ at $\lambda=0.9$ on LogDet, the accuracy drops early at $\lambda=0.3$ with erank=61.64. On the other hand, the best performance of DirectSpec$^+$ is achieved at erank=63.77, showing significant improvement over LogDet. This observation indicates that DirectSpec$^+$ can balance the spectrum in a more reasonable way, achieving a better trade-off between alignment and spectrum balancing than LogDet and AFDGCF.

\subsubsection{Comparison with Solutions to Overs-Smoothing}
As shown in Section 3, embedding collapse is attributed to the optimization part and over-smoothing (for GNN-based methods). In this subsection, we investigate (\romannum{1}): if the solutions to over-smoothing can also well address the collapse issue and (\romannum{2}): does DirectSpec$^+$ show superiority over them in addressing the over-smoothing. To this end, we compare DropEdge  \cite{rong2020dropedge} (implemented on LightGCN), a solution to over-smoothing, LightGCN (suffer from both (\romannum{1}) and (\romannum{2})), and BPR (suffer from (\romannum{1})) with DirectSpec$^+$. We report how their eranks change throughout the training in Figure \ref{other_methods_compare} (b). We can see the erank of DropEdge is higher than LightGCN while lower than BPR during the training, indicating that the solution to over-smoothing can only alleviate (\romannum{2}) while fails to tackle (\romannum{1}), the main cause of collapse issue analyzed in this work. Furthermore, DirectSpec$^+$ shows a higher and more stable erank than DropEdge, demonstrating a better capability in addressing over-smoothing than DropEdge.

\begin{figure} \centering 
\subfigure[CiteULike] {  
\includegraphics[width=0.42\columnwidth]{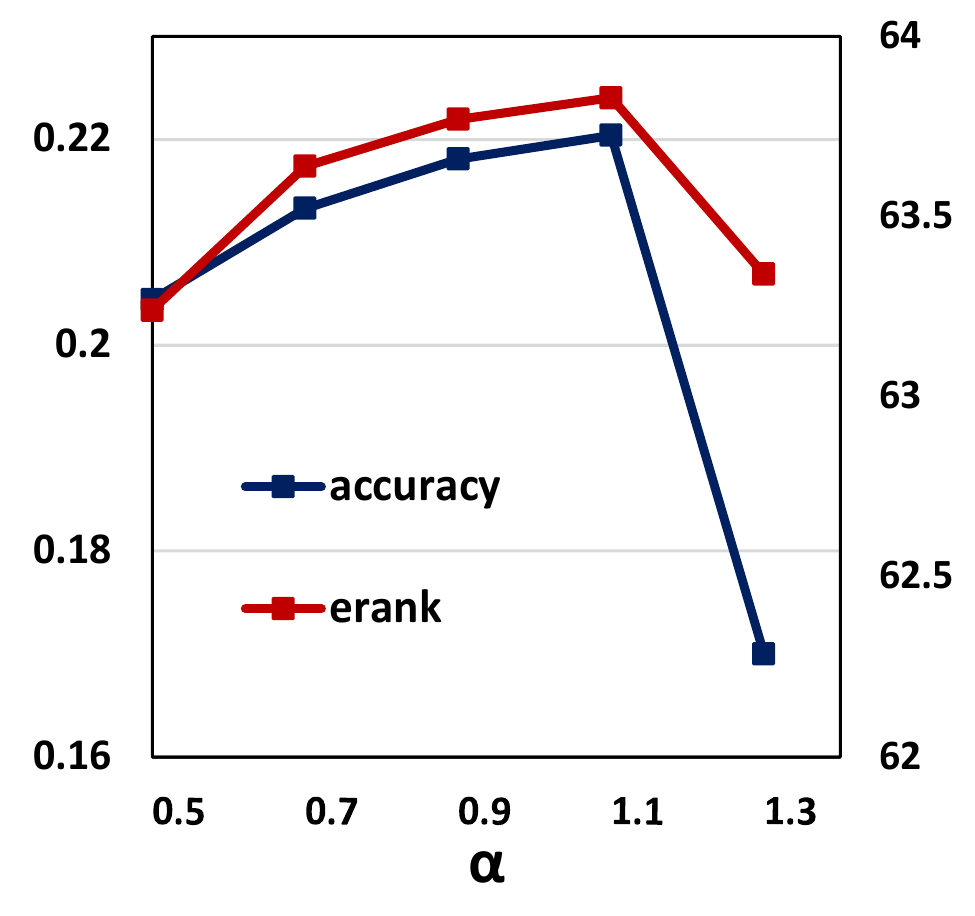} 
}
\hspace{0.2cm}
\subfigure[Yelp] {  
\includegraphics[width=0.42\columnwidth]{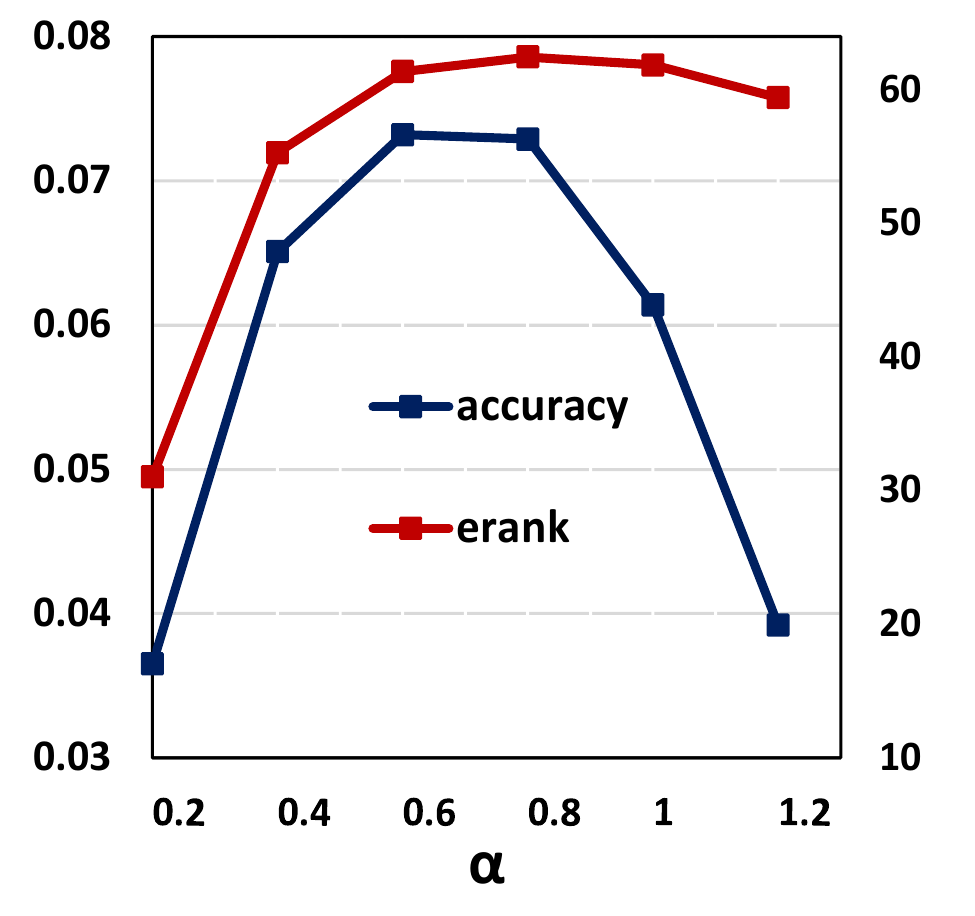} 
}
\caption{Impact of $\alpha$ on accuracy (nDCG@10) and erank.}           
\label{alpha}
\end{figure}

\begin{figure} \centering 
\subfigure[CiteULike] {  
\includegraphics[width=0.42\columnwidth]{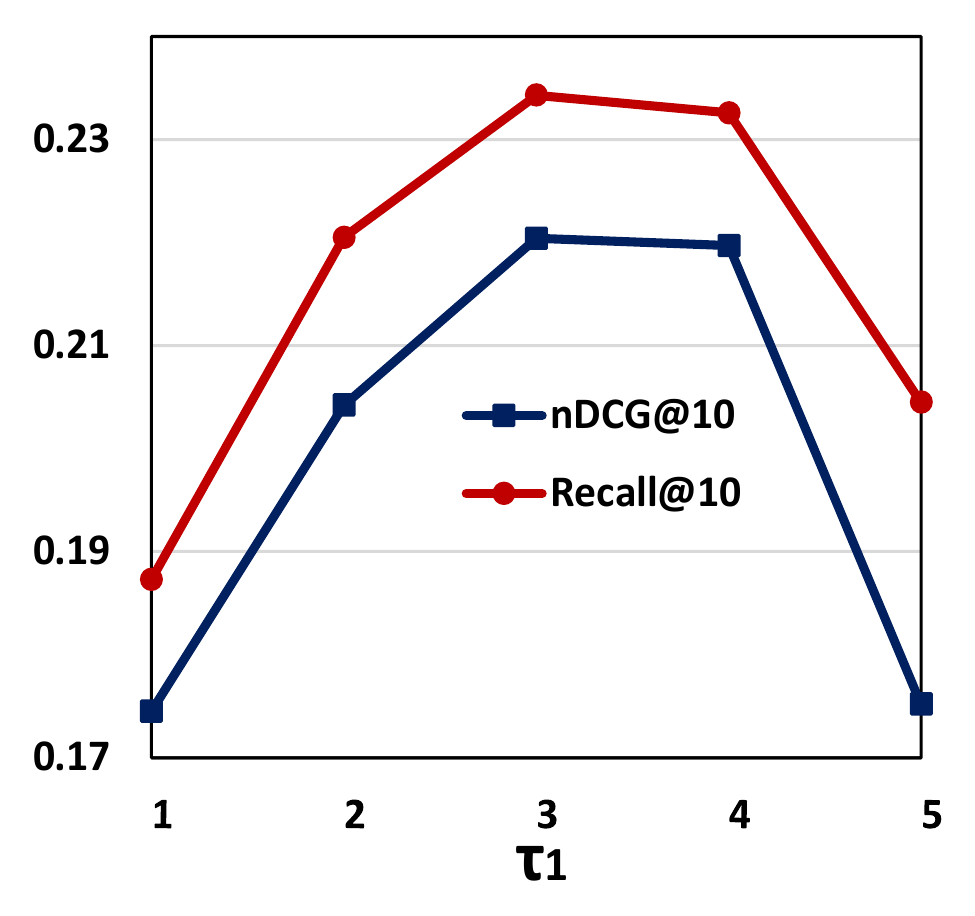} 
}
\hspace{0.2cm}
\subfigure[Yelp] {  
\includegraphics[width=0.42\columnwidth]{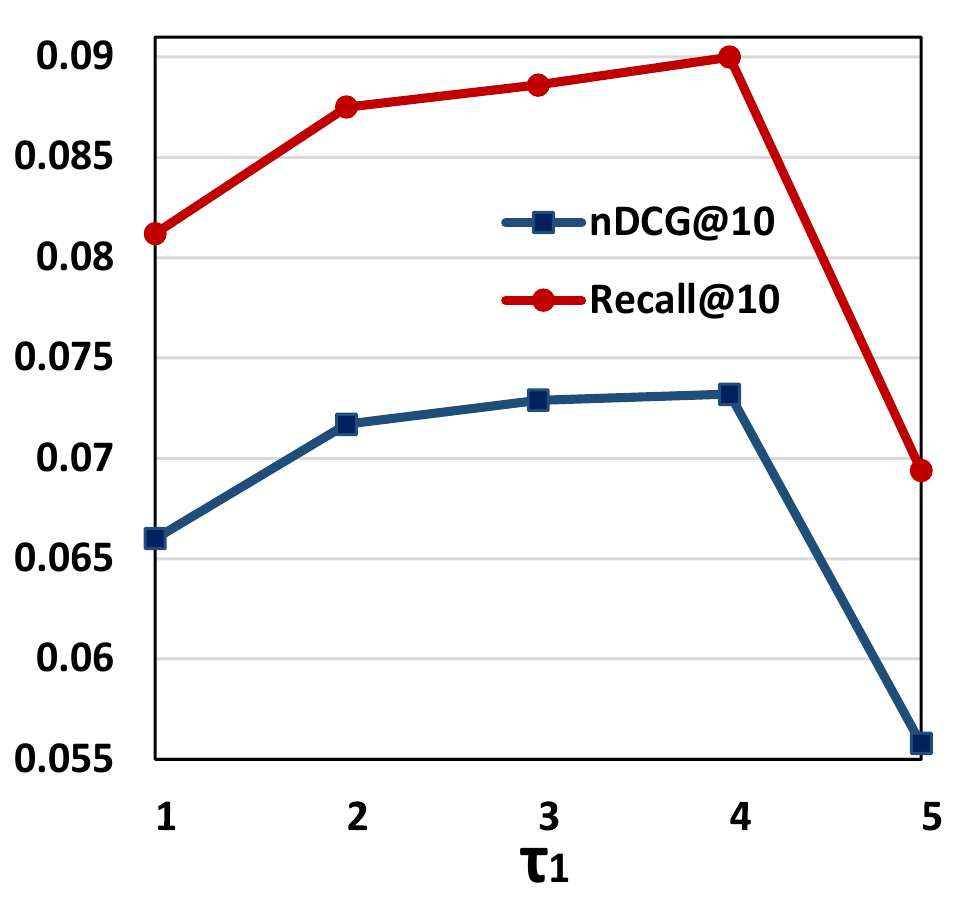} 
}
\hspace{0.2cm}
\subfigure[Gwoalla] {  
\includegraphics[width=0.42\columnwidth]{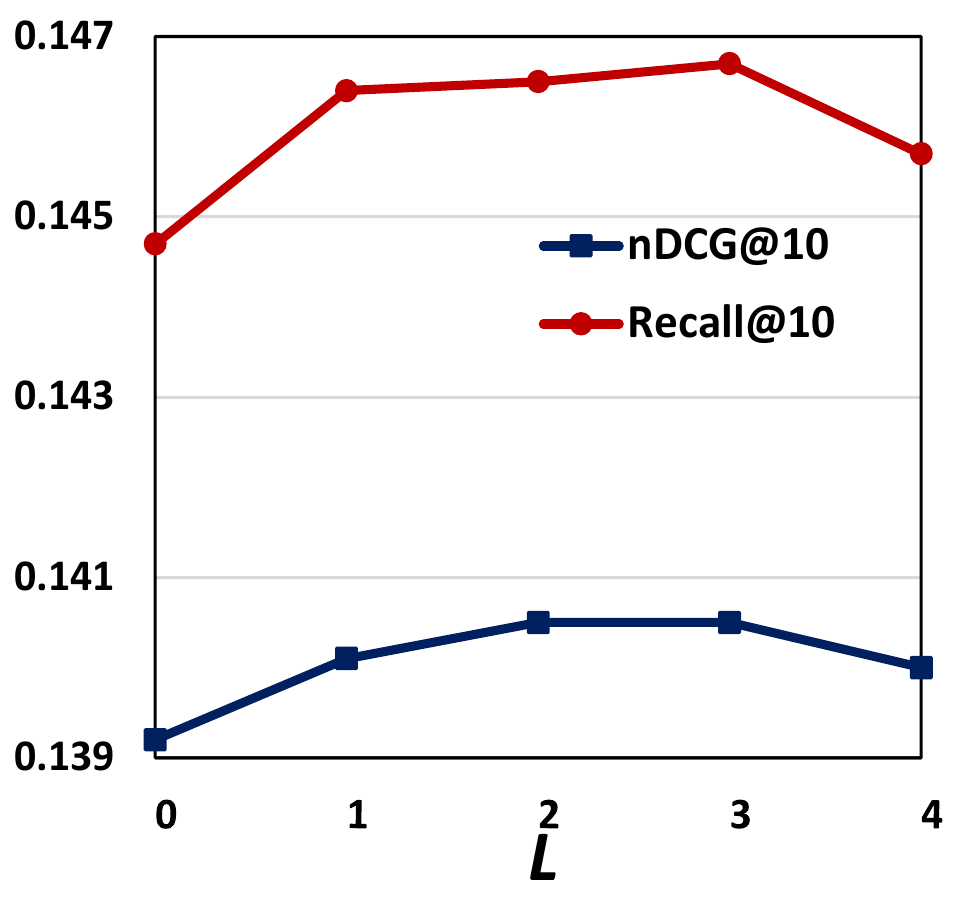} 
}
\hspace{0.2cm}
\subfigure[Gowalla] {  
\includegraphics[width=0.42\columnwidth]{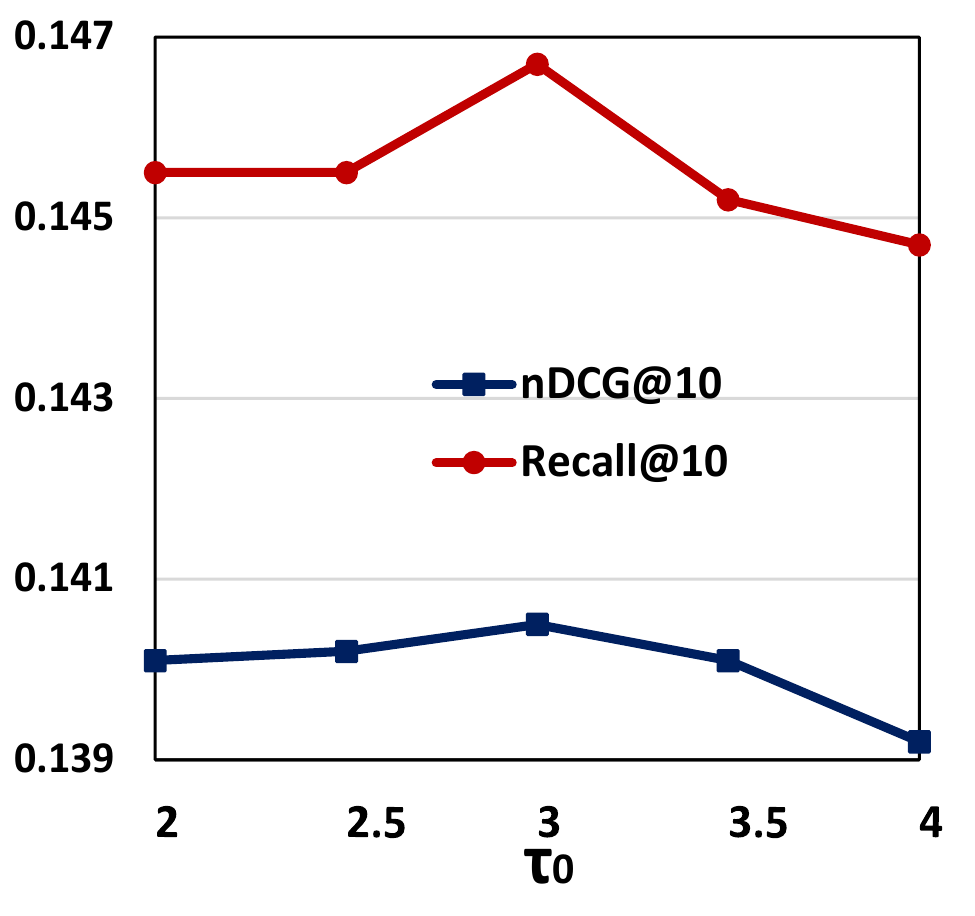} 
}
\caption{(a) and (b) show how the accuracy changes with $\tau_1$, (c) displays the sensitivity of $K$, and (d) illustrates how the accuracy changes with $\tau_0$.  }           
\label{tau}
\end{figure}

\subsection{Model Analysis}
\subsubsection{Can DirectSpec Prevent Collapse?}
In Figure \ref{prevent_collapse} (a), we report how the erank changes as training proceeds on BPR, LightGCN, and DirectSpec$^+$(MF). For BPR and LightGCN, the erank plunges significantly in the first place, then increases slowly and tends to converge to a value smaller than $d$, resulting in an incomplete embedding collapse. While the erank of DirectSpec$^+$(MF) is close to $d$ at the beginning of training and remains almost unchanged throughout the training. Figure \ref{prevent_collapse} (b) illustrates the normalized singular value distribution of the embedding matrix when the accuracy of BPR and DirectSpec$^+$(MF) is maximized. We can see that most singular values of BPR's embeddings are less than 0.2, barely contributing to the representations; while DirectSpec$^+$(MF)'s singular values are mostly larger than 0.5, indicating that the representations are contributed by more dimensions.

\subsubsection{DirectSpec vs. DirectSpec$^+$}
The goal of DirectSpec$^+$ is not simply to generate a perfectly balanced spectrum like DirectSpec, but to balance the spectrum while also preserving the mapped relationships among users/items in the embedding space. In Figure \ref{prevent_collapse} (c), the average similarity on pairs in test sets are defined as: $SIM=\frac{\sum_u \frac{\sum_{i\in \mathcal{N}^t_u} \sigma(\mathbf{h}_u^T \mathbf{h}_i)}{|\mathcal{N}^t_u|}}{|\mathcal{U}|}$ where $\mathcal{N}^t_u$ is the item set $u$ has interacted with in test sets and $\sigma(\cdot)$ is the sigmoid function. In spite of a higher erank DirectSpec has shown during training, DirectSpec$^+$ still outperforms DirectSpec as it shows a higher average similarity on test sets, indicating that DirectSpec$^+$ is better at representing the user/item relationships in the embedding space in a reasonable way.

\begin{table}[]
\caption{Comparison between static and dynamic temperature design.}
\scalebox{0.92}{
\begin{tabular}{ccccccc}
\toprule
               & \multicolumn{2}{c}{\textbf{Citeulike}} & \multicolumn{2}{c}{\textbf{Yelp}}     & \multicolumn{2}{c}{\textbf{Gowalla}}  \\ \hline
               & \textbf{nDCG@10}  & \textbf{Recall@10} & \textbf{nDCG@10} & \textbf{Recall@10} & \textbf{nDCG@10} & \textbf{Recall@10} \\
Static (i)   & 0.2197            & 0.2352             & 0.0743           & 0.0909             & 0.1389           & 0.1447             \\
Dynamic (ii) & 0.1593            & 0.1693             & 0.0485           & 0.0611             & 0.1126           & 0.1144             \\ \bottomrule
\end{tabular}}
\label{static_dynamic_compare}
\end{table} 

\subsubsection{Impact of $\alpha$}
$\alpha$ serves as a global coefficient controlling the extent of spectrum balancing. A larger (smaller) $\alpha$ implies that large singular values are multiplied by smaller (larger) weights, leading to a more (less) uniform spectrum distribution. According to Equation (\ref{alg1_svd}), DirectSpec could fail when $\alpha$ is set too large: $1-\alpha\sigma_i^2<0$ (with $K=1$). Since it takes additional time to calculate $\sigma_1$ during each training, we use the grid search for optimal $\alpha$. As shown in Figure \ref{alpha}, both the accuracy and erank first increase and then drop, showing a similar trend. Specifically, the speed of collapse tends to outrun the speed of spectrum balancing when $\alpha$ is small as erank converges to a value much smaller than $d$. While when $\alpha$ is large, the large singular values in the original distribution are multiplied by too small weights, which might cause a sharp singular value distribution similar to the original one, explaining why erank declines as consistently increasing $\alpha$. In addition, the best performance is achieved at $\alpha=1.1$ on CiteULike and $\alpha=0.6$ on Yelp, and the erank and accuracy are more sensitive when $\alpha$ is relatively small on Yelp, which might be related to the fact that Yelp suffers more from the collapse issue than other datasets.    

\subsubsection{Adaptive Temperature Design}
We first compare the dynamic and static designs proposed in Section 4.3, and report the results in Table \ref{static_dynamic_compare}. We observe that the static design outperforms the dynamic design by a large margin on all datasets, and the reason might be attributed to the data scarcity on CF task, making the algorithm difficult to learn the inherent user/item relations from the interactions data in a dynamic way. Therefore, DirectSpec$^+$ is implemented with a static temperature design, which avoids bringing additional complexity.\par

For the static temperature design, we first fixed $\tau_0=\tau_1$ and only tune $\tau_1$, then we gradually reduce $\tau_0<\tau_1$ and observe how accuracy changes accordingly. As shown in Figure \ref{tau} (a) and (b), the accuracy first increases and then drops as $\tau_1$ increases, which is maximized at $\tau_1=3$ on CiteULike and $\tau_1=4$ on Yelp. $\tau_1$ controls the strength of penalties on the pairs severely suffering from the collapse issue (\textit{i.e.,} highly correlated), the samples that are not highly correlated tend to be ignored when $\tau_1$ is set too large, while a too small $\tau_1$ fails to optimize the highly correlated pairs in an effective way (DirectSpec$^+$ degenerates to DirectSpec when $\tau_1\rightarrow0$). We take the relations among unobserved pairs into consideration via a polynomial graph filter $\sum_{l=0}^L \mathbf{\hat{A}}^l$ as reducing $\tau_0<\tau_1$ (\textit{i.e.,} the pairs showing closer relations are optimized with a lower intensity$<\tau_1$). In Figure \ref{tau} (c), DirectSpec$^+$ achieves the best performance at $L=3$, incorporating neighborhood farther than $L>3$ results in degradation as they are too far away on the graph. In Figure \ref{tau} (d), the accuracy increases and then drops as reducing $\tau$, the maximum is reached at $\tau=3.0$ on Gowalla, demonstrating it effectiveness. We also observe that the improvement from adjusting $\tau_0$ is less significant on the other two datasets, which might be related to the data density, as users/items are less hierarchically distributed on denser data. Therefore, we can simple fix: $\tau_0=\tau_1$ for such datasets.

\subsubsection{What Factors Affect DirectSpec's Performance}
As analyzed previously, DirectSpec's performance is primarily influenced by the extent of collapse and the choice of encoders (\textit{i.e.,} the algorithm implemented on). The variation in improvement caused by different degrees of collapse is unavoidable as our goal is to eliminate the collapse issue. We can choose the encoder according to the data density: GNN-based methods perform better on sparse data while MF-based methods excel in dense data.

\section{Related Work}
\subsection{Collaborative Filtering}
Collaborative Filtering (CF), a fundamental task for recommender systems, makes predictions based on user's past behaviours. Early CF methods exploit users that share similar interests or items that tend to be interacted by similar users to infer user preference \cite{sarwar2001item}. Nowadays, model-based methods are prevalent in CF and recommender systems. The core idea is to map the high dimensional sparse recommendation data to a low-dimensional space, where users and items are characterized as learnable vectors and are optimized based on observed interactions \cite{koren2009matrix}. With the development of computer hardware and communication network, more information can be collected to analyze user behaviour with advanced algorithms. Besides user-item interactions, side information such as social relations \cite{jiang2012social}, reviews \cite{bao2014topicmf}, geological information \cite{lian2014geomf,peng2018vector}, and so on can help the algorithms better understand user taste and preference. On another line, early simple CF methods \cite{koren2009matrix,ning2011slim} are replaced by more powerful algorithms such as neural network \cite{sedhain2015autorec,he2017neural}, attention mechanism \cite{chen2017attentive,kang2018self}, transformer \cite{sun2019bert4rec}, etc. Recently, graph neural network (GNN) has also shown tremendous success in recommender systems \cite{wang2019neural,peng2022svd,peng2024powerful}. By representing interactions as a bipartite graph, the core idea of GNN is to exploit higher-order neighbor connections to facilitate user/item representations \cite{kipf2017semi}. However, it has been reported that multi-layer GNNs suffer from an over-smoothing issue that causes user/item representations to be indistinguishable \cite{peng2022less,li2018deeper}. We showed that an embedding collapse issue resembling the over-smoothing exists on recommendation algorithms in general, and our proposed approach has also been shown effective tackling over-smoothing.   

\subsection{Collapse in Representation Learning} 
Representation learning has attracted tremendous attention in various research fields \cite{mikolov2013efficient,hjelm2018learning}, the goal of which is to represent data in an effective and efficient way to make it easier to extract useful information when building predictors. However, it has also been shown that the model tends to collapse where all inputs are mapped to the same constant vector, when only optimizing the model based on the positive pairs \cite{chen2021exploring}. This issue can be well alleviated by self-supervised learning and contrastive learning by exploiting negative samples in an effective way \cite{hjelm2018learning,chen2020simple}. Due to the heavy computation cost, some research effort has been made to simplify the algorithms without explicitly sampling negative data \cite{chen2021exploring,grill2020bootstrap,zbontar2021barlow,zhang2023mitigating}. Unfortunately, subsequent works show that a dimensional collapse cannot be ignored where the embedding vectors end up spanning a lower dimensional subspace of the whole embedding space \cite{hua2021feature,jing2021understanding}. The collapse issue in representation learning shares similarities to some issues such as over-smoothing in GNN \cite{li2018deeper}, which inspires some researchers to tackle issues in GNN \cite{guo2023contranorm}. \par

Inspired by the aforementioned works tackling collapse in representation learning, we reviewed existing recommendation methods, and showed that they suffer from a collapse issue as well. Particularly, the alignment of positive pairs results in complete collapse, and an incomplete collapse still exists despite introducing negative samples~\cite{rendle2009bpr,he2017neural} where the representations are predominantly distributed in certain dimensions. We address this issue from a spectral perspective by directly optimizing the spectrum distribution, ensuring that users/items span the whole embedding space effectively. Moreover, we establish the connection between our proposed DirectSpec and SSL methods without explicit negative sampling by showing they they can be considering as special cases of DirectSpec. Lastly, some recent works   \cite{chen2023towards,zhang2023mitigating} pointed out a dimensional collapse issue for CF sharing similarity with our work. Since (\romannum{1}): the codes have not been officially released and (\romannum{2}): the proposed methods shares similarities ( based on the LogDet function), we implemented \cite{zhang2023mitigating} as the baseline, and the performance comparison demonstrated the superiority of our proposed methods over them in terms of accuracy and efficiency.

\section{Conclusion and Limitation}
In this work, we showed that existing CF methods mostly suffer from an embedding collapse issue. Particularly, alignment of positive pairs is equivalent to a low pass filter, causing the representations of users and items to collapse to a constant vector. Exploiting negative pairs (\textit{i.e.,} negative sampling) can decelerate collapse by acting as a high pass filter to balance the embedding spectrum, while the effectiveness is only limited to certain loss functions which still results in an incomplete collapse, where embeddings are distributed along certain dimensions instead of making full use of all dimensions. To tackle this issue, we proposed a DirectSpec which directly optimizes the spectrum distribution to ensure that all dimensions can contribute to the user/item embeddings as equally as possible. We further pointed out that the key of high-quality representation lies in trade-off between alignment and spectrum balancing, and proposed an enhanced algorithm DirectSpec$^+$, which can balance the spectrum more adaptively and individually. Moreover, we established the connection with SSL-based methods without explicit negative sampling such as Barlow-Twin, LogDet, and spectral contrastive loss by showing that they can be considered as special cases of DirectSpec. Finally, extensive experiments on three publicly available datasets demonstrated the effectiveness and efficiency of our proposed methods. Given the generality of our DirectSpec learning paradigm, we believe that our proposed method can effectively address collapse in representation learning more broadly.\par

We acknowledge that our proposed methods are still far from perfect, despite the efficiency and effectiveness. Firstly, balancing the embedding spectrum requires tuning hyperparameters such as $\alpha$ and $\tau$, reducing the utility. Furthermore, the effectiveness of some designs (\textit{e.g.,} $\tau_0$) is not significant and varies on datasets. We aim to address above limitations in the future work.

\noindent
\textbf{Acknowledgement}\\
This paper is based on results obtained from the project, “Research and Development Project of the Enhanced Infrastructures for Post-5G Information and Communication Systems” (JPNP20017), commissioned by the New Energy and Industrial Technology Development Organization (NEDO).

\bibliographystyle{ACM-Reference-Format}
\bibliography{sample-base}
\end{document}